\documentclass[trackchanges,twocolumn,tighten]{aastex62} 

\usepackage{amsmath}

\usepackage{longtable}
\usepackage{soul}
\usepackage{CJK}
\usepackage{lineno}

\newcommand{\Msun}{M_\odot}
\newcommand{\Rsun}{R_\odot}
\newcommand{\Lsun}{L_\odot}
\newcommand{\vc}{v_c}

\newcommand{\grad}{\nabla}

\newcommand{\grade}{\nabla_\mathrm{e}}

\newcommand{\Teff}{T_\mathrm{eff}}

\newcommand{\bbud}{\alpha\;\mathrm{Ori~B}}

\definecolor{meridithgreen}{RGB}{0, 150, 0}
\definecolor{jaredpurple}{RGB}{93, 63, 211}

\newcommand{\gtapprox}{\gtrsim}
\newcommand{\ltapprox}{\lesssim}

\newcommand{\appropto}{\mathrel{\vcenter{
		\offinterlineskip\halign{\hfil$##$\cr
	\propto\cr\noalign{\kern2pt}\sim\cr\noalign{\kern-2pt}}}}}

\newlength{\apjcolwidth}
\newlength{\figwidth}
\newlength{\doublewide}
\begin{document}
\title{A Buddy for Betelgeuse: Binarity as the Origin of the Long Secondary Period in $\alpha$~Orionis} 

\author[0000-0003-1012-3031]{Jared A. Goldberg}
\affiliation{Center for Computational Astrophysics, Flatiron Institute, New York, NY, USA}

\author[0000-0002-8717-127X]{Meridith Joyce}
\affiliation{University of Wyoming, 1000 E University Ave, Laramie, WY USA}
\affiliation{Konkoly Observatory, HUN-REN CSFK, Konkoly-Thege Mikl\'os \'ut 15-17, H-1121, Budapest, Hungary}
\affiliation{CSFK, MTA Centre of Excellence, Konkoly-Thege Mikl\'os \'ut 15-17, H-1121, Budapest, Hungary}

\author[0000-0002-8159-1599]{L\'{a}szl\'{o} Moln\'{a}r}
\affiliation{Konkoly Observatory, HUN-REN CSFK, Konkoly-Thege Mikl\'os \'ut 15-17, H-1121, Budapest, Hungary}
\affiliation{CSFK, MTA Centre of Excellence, Konkoly-Thege Mikl\'os \'ut 15-17, H-1121, Budapest, Hungary}
\affiliation{E\"otv\"os Lor\'and University, Institute of Physics and Astronomy, P\'azm\'any P\'eter s\'et\'any 1/A, H-1117, Budapest, Hungary}

\correspondingauthor{J.~A.~Goldberg}
\email{jgoldberg@flatironinstitute.org}

\begin{abstract}
We predict the existence of $\alpha$~Ori~B, a low-mass companion orbiting Betelgeuse. This is motivated by the presence of a 2170-day Long Secondary Period (LSP) in Betelgeuse's lightcurve, a periodicity $\approx5$ times longer than the star's 416 day fundamental radial pulsation mode. While binarity is currently the leading hypothesis for LSPs in general, the LSP and the radial velocity variation observed in Betelgeuse, taken together, necessitate a revision of the prevailing physical picture. Specifically, the lightcurve-RV phase difference requires a companion to be \textit{behind} Betelgeuse at the LSP luminosity minimum, $\approx$180 degrees out of phase with the system orientation associated with occultation. We demonstrate the consistency of this model with available observational constraints and identify tensions in all other proposed LSP hypotheses. Within this framework, we calculate a mass for $\alpha$~Ori~B of $M\sin i=1.17\pm0.7\,M_\odot$ and an orbital separation of $1850\pm70\,R_\odot$, or $2.43^{+0.21}_{-0.32}$ times the radius of Betelgeuse. We then describe the features of the companion as constrained by the fundamental parameters of Betelgeuse and its orbital system, and discuss what would be required to confirm the companion's existence observationally. 
\end{abstract}

\keywords{Betelgeuse, Long Secondary Periods, Variability, Red Supergiants, Binarity}

%
%
\section{Introduction}
\label{sec:intro}
Interest in Betelgeuse, more formally known as $\alpha$ Orionis, has blossomed over the past five years due in large part to its anomalously severe dimming observed between the end of 2019 and beginning of 2020 \citep{Guinan2019, Harper2020, Dupree2020}. 
Now widely understood to have been caused by a dust cloud \citep{Montarges2021}, the ``Great Dimming'' nonetheless inspired many groundbreaking investigations into our nearest supergiant. These studies have led to revisions in our understanding of Betelgeuse's behavior and fundamental parameters, as well as opened new lines of inquiry into some of Betelgeuse's less well-understood properties, such as its apparent 5 km/s surface rotation rate \citep{Ma2024}, the characterization of its variability \citep{Jadlovsky2024}, and the classification of its pulsation modes \citep{Joyce2020, Saio2023, MacLeod2023}.  

A distinctive variability cycle observed in Betelgeuse is a $\approx$2100 day period in the lightcurve, \citep[see, e.g.][]{Kiss2006,Joyce2020,Granzer2022,MacLeod2023}, often labelled a Long Secondary Period (LSP). LSPs are seen in many cool, luminous stars ranging from Red Giant Branch stars (RGB; e.g.\ \citealt{Wood1999}), Asymptotic Giant Branch stars (AGB; e.g.\ \citealt{Zijlstra2002, Uttenthaler2015}), post-AGB stars \citep{vanWinckel1999,Bodi2016}, Yellow Hypergiants (YHGs; e.g. \citealt{PercyKim2014}), and Red Supergiant (RSG; e.g. \citealt{Kiss2006}) stars, in all cases showing generally steady periods ranging from hundreds to thousands of days (e.g.\ \citealt{Takayama2023}). 
Long Secondary Periods are an as-yet unexplained class of stellar variability, called `secondary' because LSPs are typically a few to ten times slower than their host stars' radial fundamental pulsation modes (FMs).

Outstanding among the questions raised by renewed fascination with and deeper investigation into Betelgeuse is whether Betelgeuse's prominent, $\approx$2100-day periodicity is an LSP or the FM. 
The answer to this question carries significant implications for Betelgeuse's current evolutionary phase, which in turn implies the timeline for Betelgeuse's death in a supernova explosion \citep{Wheeler2017,Nance2018,Chatzopoulos2020,Saio2023}. If the $\approx$2100-day periodicity is the fundamental mode, it implies a large and observationally contentious radius for Betelgeuse. Further, it would place Betelgeuse's current evolutionary stage beyond the onset of core carbon burning, suggesting that a supernova explosion is imminent within the next several dozen to several hundred years \citep{Saio2023}. If, on the other hand, the 2100-d periodicity is an LSP, Betelgeuse is comfortably within in its core helium burning phase and not due for an explosion for hundreds of thousands of years \citep{Joyce2020}. As we discuss below, most studies directly or indirectly support classifying the $\approx$420-day periodicity 
as the fundamental mode. As such, the cause of Betelgeuse's 2100-day periodicity remains elusive. 

There have been several mechanisms proposed to explain LSPs in general, ranging from giant cell convection, mode interactions, non-radial pulsations, and binarity, as well as more exotic physics, such as dust modulation beyond the stellar surface (many of these are discussed in \citealt{Wood2004IAU}). We explore the relative merits of each of these and more hypotheses proposed in the literature
individually in the context of modern observational data (see Section \ref{sec:hypotheses}). We ultimately identify binarity as the most likely among them.

The idea that binarity could explain LSPs dates back to the late 1990s and early 2000s \citep{Wood1999, Wood2004} and has been reinforced since by the sequences formed by variable stars on period-luminosity diagrams \citep{Soszynski2007} and in RV data \citep{Hinkle2002,Nicholls2009}. 
While such early versions of the binarity hypothesis were most concerned with whether a close companion could induce low-frequency modes on the primary star, the current leading theory is that the timescale of the LSP period is set by the orbital time of a low-mass companion, and the mechanism of dimming and brightening involves the formation and removal of dust along the line of sight in phase with the companion's orbit \citep{Soszynski2021}. Additionally, it has been proposed before that Betelgeuse is a multiple system \citep{Karovska1986} from speckle-imaging measurements, inferred to be two high-eccentricity companions including one at a separation of 0.06" with a period of 2.1 years, which has not been confirmed in follow-up work with higher-resolution instruments \citep[e.g.][]{Wilson1992, Kervella2009,Montarges2016}.

In this paper, we explore the implications and requirements of the binarity-as-LSP model for $\alpha$~Orionis. In Section \ref{sec:LSPs}, we discuss LSPs in cool, evolved stars in the context of the period-luminosity sequences formed by classical variable stars and consider where Betelgeuse fits in to this picture.
In Section \ref{sec:phaseoffset}, we compare the brightness and radial velocity variations observed in Betelgeuse and 
a significant subset of \textit{Gaia} LSP stars
and discuss the physical implications of the LC-RV phase offsets. In Section \ref{sec:hypotheses}, we overview all of the scenarios proposed as an explanation for Betelgeuse's LSP, demonstrating critical flaws in all but one case: Betelgeuse has a companion that interacts with the star's dusty circumstellar environment. In Section \ref{sec:properties}, we derive the properties this companion must have based on observational constraints (\ref{subsec:prop}) and discuss the conditions under which it should be observable (Section~\ref{subsec:observability}). We summarize and present our conclusions
in Section \ref{sec:conclusions}.

%
%
\section{LSPs in Cool Evolved Stars} 
\label{sec:LSPs}

LSPs, sometimes referred to as $P_2$ in older literature, are a common but not ubiquitous phenomenon in cool evolved stars. They have been observed in $\approx30\%$ of long-period variables \citep[LPVs; see, e.g.][]{Soszynski2021,Pawlak2024} and have been detected both in the Milky Way and in the Magellanic Clouds \citep[see, e.g.][]{Payne1954,Wood1999,Kiss2006,soszynski2009-lmc,soszynski2011-smc,soszynski2013-bulge}. These stars have large, radially extended convective envelopes, and with increasing luminosity, the convection in their envelopes can become near- or trans-sonic, with large-scale convective plumes spanning large fractions of the stellar radius \citep[see, e.g., review by][]{Chiavassa2024}. Stars hosting Long \textit{Secondary} Periods
by definition show other pulsation modes, which are identified as low-order radial pressure modes (e.g.\, fundamental and first-overtone, or FM and O1). 

\begin{figure}[]
\centering
\includegraphics[width=0.49\textwidth]{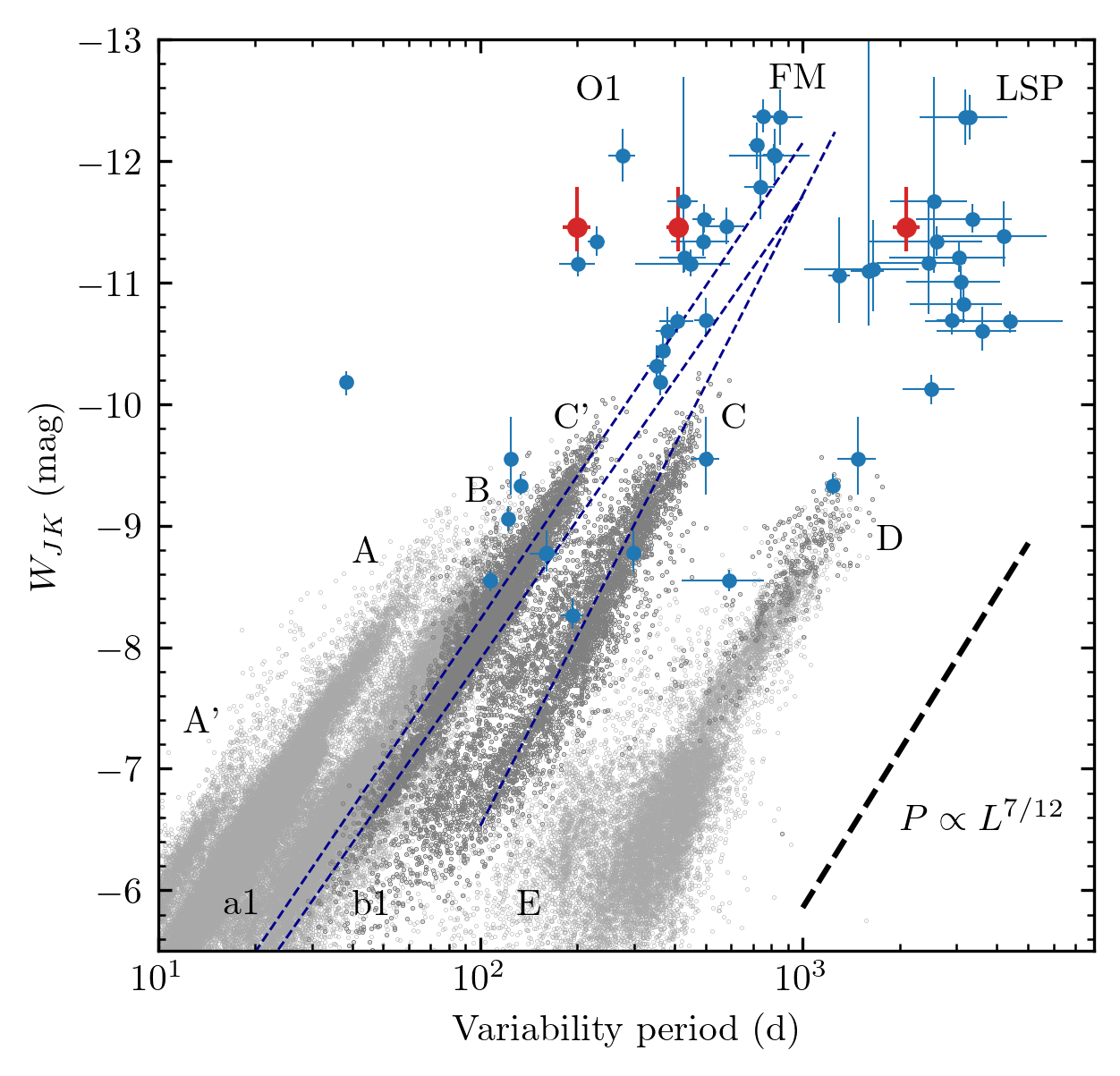}
\caption{
Period-Wesenheit luminosity relations for the OGLE-III long-period variables in the LMC (grey), and for Galactic red supergiants (blue) collected by \citet{Kiss2006} based on absolute $J$- and $K$-band magnitudes. Red points, from left to right, mark the positions of the O1, FM and LSP periodicities for Betelgeuse. Darker grey points are semiregular and Mira variables pulsating in the FM and O1 modes. Thin dashed lines indicate various PW relations for RGB, AGB and Mira stars in the LMC from \citet{soszynski-2007}.  
}
\label{fig:PL}
\end{figure}

Large sky surveys targeting the Magellanic Clouds, such as the MACHO and OGLE surveys \citep{MACHO-1997,Udalski-1992,Udalski-2003}, revealed that variability in red giant stars fall onto multiple, nearly parallel period-luminosity (P-L) relations, which were originally labeled from A to D 
\citep{Wood1999}. Sequences A, B and C were attributed to pulsation modes, whereas a fifth sequence, E, was attributed to eclipsing binaries 
\citep{Derekas2006}. Sequence D included the Long Secondary Periods, which were found to be too long to be radial pulsation modes. We reproduce the sequences for the LMC from the OGLE-III survey in Figure~\ref{fig:PL}, which includes the later-identified A' and C' sequences \citep{soszynski2009-lmc}. Here we plot the absolute K magnitudes of the stars, after cross-matching the OGLE catalog with the IRSF Magellanic Clouds Point Source Catalog \citep{Kato-2007} using a distance modulus of $\mu_\text{LMC} = 18.54$~mag \citep{wielgorski2022}.

Classical pulsating stars have been known to form P-L relations thanks to the pioneering work of \citet{Leavitt-1912}. These extend from classical Cepheids to other classes, such as Type II Cepheids, RR~Lyrae and $\delta$~Scuti stars, but also to red giants,
semiregular and Mira stars \citep{wyrzy-2012}. Many authors, e.g., \citet{Yu2020} and \citet{Trabucchi2017}, have studied how the red giant sequences relate to specific modes. The latter work concluded that sequence A', A and B correspond to the third, second and first radial overtones, respectively, while C' and C separate less clearly due to the combined presence of RGB and AGB stars. C' includes both overtone and fundamental-mode stars, whereas C is largely composed of Mira stars, which are AGB stars pulsating in the fundamental mode (e.g.\ \citealt{Molnar2019, Joyce2024}). 

LSPs in sequence D also follow a period-luminosity relationship that is distinct from the P-L relations discussed so far.
Sequence D was discovered by \citet{Wood1999} using MACHO data. Binarity was proposed as a possible mechanism for the LSP by various authors \citep{Hinkle2002,Olivier2003,Wood2004}. This was reaffirmed 
when \citet{Derekas2006} showed that sequence D is an extension of sequence E (eclipsing binaries).

These sequences can also be extended further into the RSG regime (also shown in Figure~\ref{fig:PL}). \citet{Kiss2006} found that Galactic supergiants form sequences that correspond to the O1, FM and LSP periodicities and extend from sequences B, C, and D, respectively.
We recalculated the $M_J$ and $M_K$ absolute brightnesses of this RSG sample, using \textit{Gaia} DR3 distances from the catalog of \citet{bailer-jones-2021} and plot these as blue filled circles in Figure~\ref{fig:PL}. These distances were sufficient in most cases, but a few very bright stars required other sources for parallax \citep{hipparcos-2007,Reid-2019}. For Betelgeuse, we used the distance calculated by \citet{Joyce2020}, shown as red filled circles. Here we show $W_{JK}$ Wesenheit indices as a proxy for luminosity based on the \textit{(J-K)} colors, using the equation: $W_{JK} = K - 0.686(J-K)$, because that limits the effects of interstellar reddening and helps to separate the sequences better \citep{lebzelter-2019}.

Compared to the original plot by \citet{Kiss2006}, the O1 and FM sequences reproduced in Figure~\ref{fig:PL} appear to be considerably tighter. 
However, mode assignment based on PL or PW relations can be complicated as the constituent stars may belong to very different evolutionary stages, luminosity regimes, and variability classes that follow different relations. To illustrate this, we include three different PW relations with blue dashed lines, computed by \citet{soszynski-2007}, who identified further sequences and substructures in the red giant sample. The PW relation running through sequence C belongs to the Mira stars, whereas a1 and b1 belong to the lower-luminosity AGB and RGB stars, respectively. The FM supergiants lie close to the extrapolations of these relations. We discuss modal assignment in more detail and based on further arguments in Sect.~\ref{subsec:notFM}.


The RSG LSP periods in Figure~\ref{fig:PL} show larger scatter and have larger uncertainties, but appear to form a coherent extension of sequence D. The periods identified in Betelgeuse as the O1, FM and LSP align with the RSG sequences. In the case of radial pressure modes, in which the pulsation period is determined by the stellar radius and the average sound speed, a P-L relation is expected, as the period scales with average density, thereby radius and mass, and thereby luminosity. In this case, the longest possible radial pressure mode is the FM. The apparent existence of a P-L relation in the LSP, on the other hand, is puzzling \textit{prima facie}. 
Gravity modes can yield periods below the FM, but no such period-luminosity relation exists for g-modes.

Unless the LSP phenomenon is simply a ubiquitous off-by-one error in the classification of the FM (see Section~\ref{subsec:notFM}), LSPs appear impossible to explain with conventional pulsation theory 
when considering only material within the stellar photosphere as the ``star.'' Whether the LSP is a pulsation or something else entirely, its period must be set by something other than the acoustic timescale of the star itself. As such, the LSP must operate on timescales longer than the dynamical timescale at the stellar surface $t_\mathrm{dyn} = R/v_\mathrm{esc}$, which is equivalent to the timescale of a Keplerian orbit at a distance equal to the stellar radius.

It is reasonable to ask why something without an intrinsic source of variability would nonetheless display a P-L relation. For this we can look outside the surface of the star, and make a dynamical argument.  Generically, for a virial fluid element or body outside the stellar radius, a period is set by the Keplerian orbital timescale at that location. Although a generic orbit is not sensitive to the stellar luminosity, any orbital architecture knows about the stellar mass and radius, as $a>R$, where $a$ is the distance from the center of the star to the external location setting the long secondary period timescale, i.e. the orbital separation. 

If $a$ scales with 
$R$, then we can estimate
\begin{equation}
P=2\pi\sqrt{\frac{a^3}{GM}}\appropto{}\frac{R^{3/2}}{M^{1/2}}\appropto{}L^{7/12},
\end{equation}
where the last proportionality makes use of the approximations that (1) $L\appropto{}M^3$ for gas-pressure-dominated stars and (2) $L\propto{}R^2T_\mathrm{eff}^4\appropto{}R^2$. This is appropriate since the surface temperature range for RSGs, RGBs, and AGBs does not vary substantially relative to $L$. Notably, the zero-point of this relation would depend on the exact relationship between $a$ and $R$, which introduces scatter in the scaling relation. This crude scaling argument nonetheless illustrates that any period sensitive to a star's radius and mass, such as one set by dynamics outside the stellar surface, is in turn sensitive to the star's luminosity.

A notable feature of observed LSPs is that they
appear more prominently in bluer bands, though the effective temperatures of their host stars are not thought to change dramatically \citep{Takayama2015,Trabucchi2023}. This suggests that LSPs are likely related to the formation, destruction, modulation, or attenuation of dust around stars. 
However, dust alone cannot explain the LSP phenomenon; the presence of a dust-modulating mechanism is also required.
When searching in visual passbands,
\citet{Soszynski2021} found that only the attenuation by the cloud was visible. However, they identified secondary eclipses at longer wavelengths from WISE photometry, which is capable of detecting infrared radiation from the dust cloud directly. 
\citet{Soszynski2021} thus conclude that LSPs arise when a small companion orbits the star together with an extended dust cloud. 

Despite what we know about LSPs observationally and what we can infer about them theoretically, they remain to date the only form of steady-state stellar variability without a strong consensus on the underlying physics. 

%
%
\section{Characterizing the Lightcurve-RV Phase Offset} 
\label{sec:phaseoffset}
An important diagnostic for the nature of pulsational versus orbital behavior is the presence or lack of radial velocity (RV) modulations and the phase relative to brightness fluctuations. 

\begin{figure}[]
\centering
\includegraphics[width=\columnwidth]{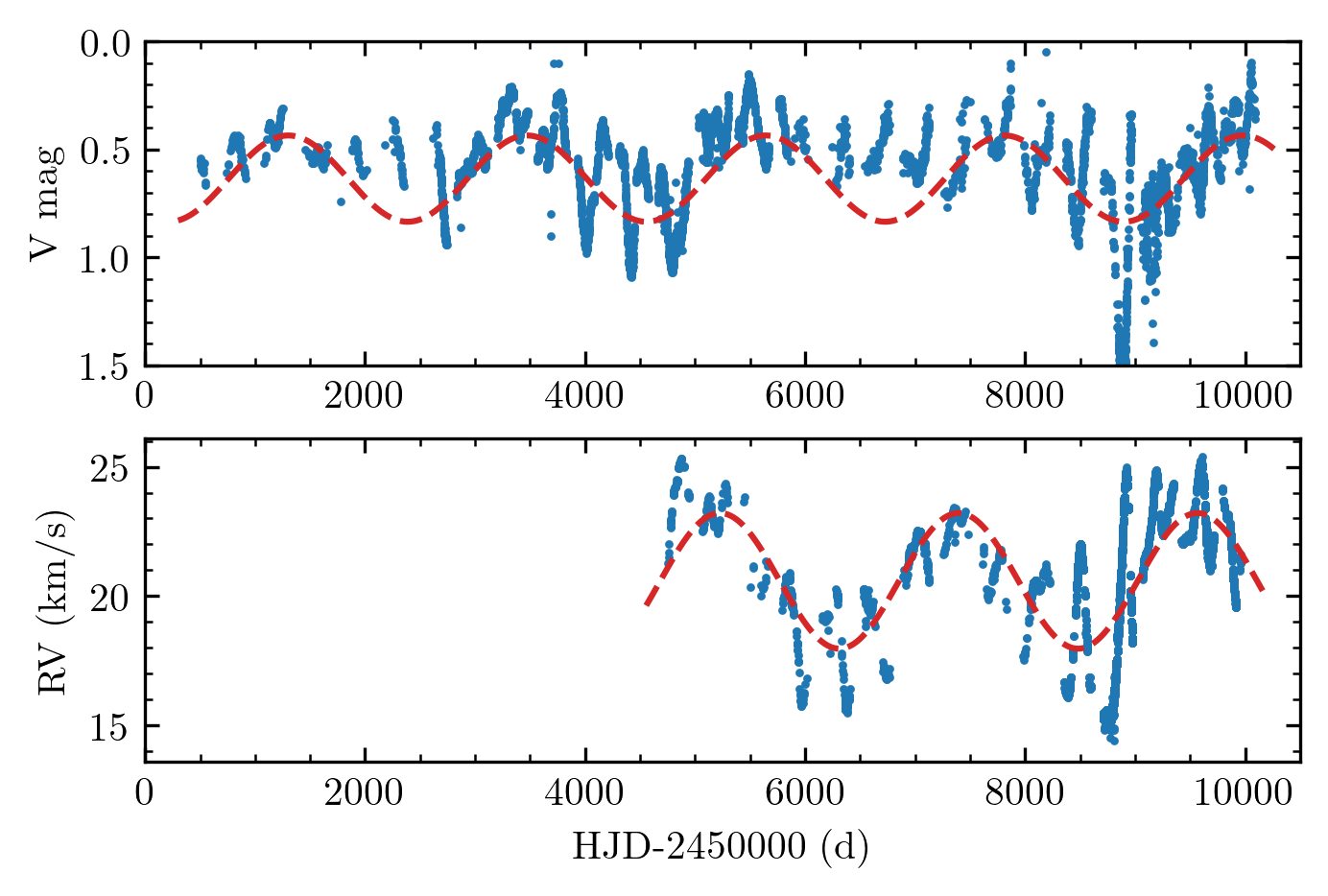 }
  \caption{We show the photometric (V-band and SMEI from \citet{Joyce2020} and from the AAVSO) and RV \citep{Granzer2022} variations for the star, plus single sine fits with red dashed lines, using the same 2170~d periodicity.
  Importantly, dimming in Betelgeuse's LSP always happens when the photosphere begins to move away from the observer, as indicated by the RV.}
     \label{fig:lightcurve_vs_RV}
\end{figure}

Figure \ref{fig:lightcurve_vs_RV} shows the recent lightcurve (LC) and RV variations for Betelgeuse side by side. The photometric data include \textit{V}-band observations collected by the American Association of Variable Star Observers (AAVSO) as well as the processed and rectified photometry of the star from the SMEI instrument \citep{Jackson-2004,Hick-2007}, originally published by \citet{Joyce2020}. RV observations were collected for more than a decade with the STELLA robotic telescope at Tenerife \citep{Granzer2022}. From the latter, an LSP period of $2169\pm5.3$~d and a peak-to-peak amplitude of $5.36\pm0.056$ km/s were determined.

We fitted both variations with the $P_{\rm LSP}=2170$~d periodicity, from which we calculate the phase offset \citep{Granzer2022,Jadlovsky2023}. This period value is consistent with other observations \citep{Kiss2006,Joyce2020}.

\begin{figure*}[]
\centering
\includegraphics[width=\textwidth]{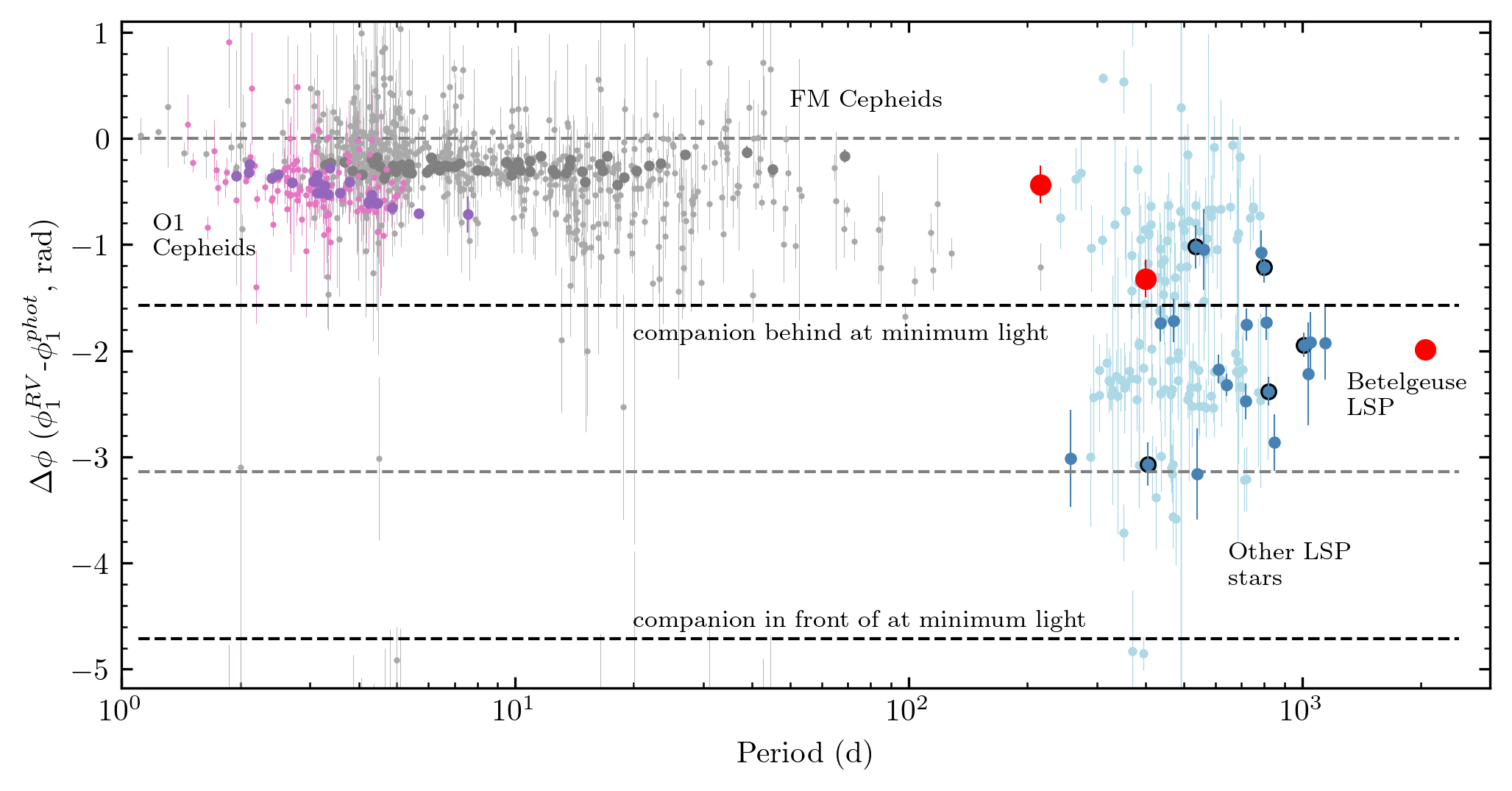}
\includegraphics[width=1.0\textwidth]{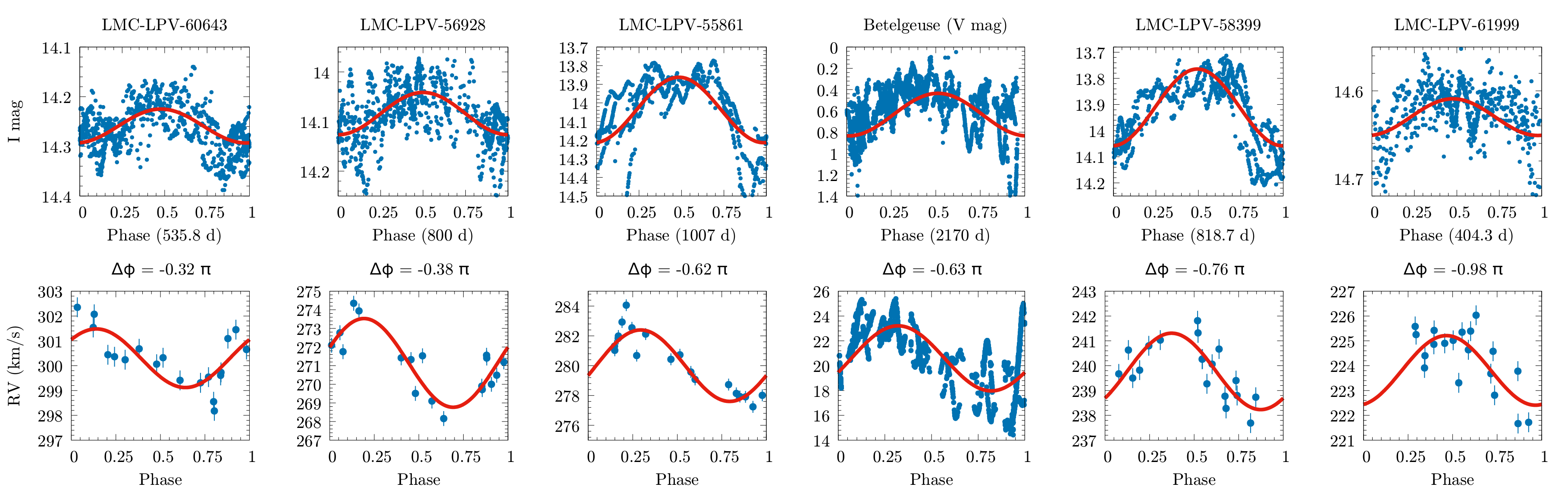}
  \caption{\textbf{Upper:} phase offset between the radial velocity and photometric variations. Classical Cepheids from Gaia DR3 \citep{Ripepi2023,Gaia-DR3-2023} are shown with pink and light gray points, while stars from \citet{Szabo2007} are shown with purple and dark grey for FM and O1 stars, respectively, for reference. LSP stars from \citet{Nicholls2009} are shown with dark blue symbols, whereas Gaia DR3 LSP candidates from \citet{Trabucchi2023} are light blue. The O1, FM and LSP points for Betelgeuse are shown in red (from left to right). \textbf{Lower:} OGLE \textit{I}-band light curves and RV curves folded with the LSP periods, including sine fits for selected stars from \citet{Nicholls2009}, ordered by decreasing phase offset, with Betelgeuse inserted into the sequence. Selected stars are marked with black outlines in the upper panel. }
     \label{fig:Gaia_LSP}
\end{figure*}

\begin{table}[]
\centering
\begin{tabular}{lccccc}
\hline
\hline
ID & \textit{P} (d) & $A_1^V$ (mag) & $A_1^{RV}$ (km/s) & $\Delta\phi_{RV-V}$ & e$\Delta\phi$ \\
\hline 
LSP & 2170  & 0.084(4) &  2.62(2) & $-1.990$ & 0.027 \\
FM  & 380 & 0.088(4) & 1.45(2) & $-1.322$ & 0.058 \\
O1 & 216 & 0.075(4) & 1.72(2) & $-0.435$ & 0.059 \\
 \hline
\hline
\end{tabular}
\caption{Fitted periods, Fourier amplitudes and phase differences for the $V$-band and RV observations of Betelgeuse we use in this paper. }\label{tab:betphase}
\end{table}

To determine the phase offset between the LC and RV for Betelgeuse, we follow the phase lag method described by \citet{Szabo2007}, where we calculate $\Delta \phi_1 = \phi^{\rm  RV}_1 - \phi_1^{\rm phot}$ from the phases of the fundamental component of the following Fourier fit to the data:
\begin{equation}
    f(t) = a_0 + \sum_{k=1}^n A_k\, \sin(2\pi f_k t + \phi_k ), 
\end{equation}
where $a_0$ is either the average brightness or the average (also called $\gamma$) velocity, $f_k$ is the common frequency between the two data sets, and $A_k$ and $\phi_k$ are the Fourier amplitude and phase components, respectively. For simplicity, we limit the fit to $k=1$. However, we calculate the phase offset in Betelgeuse not just for the LSP but for the FM and O1 modes as well, based on the $V$-band and SMEI curves from \citet{Joyce2020}, plus newer $V$-band observations collected by the AAVSO, truncated to the interval of the RV observations. We list our results for Betelgeuse in Table~\ref{tab:betphase}. The light curve and pulsation periods of Betelgeuse are known to vary, and they went through considerable changes after the Great Dimming again, with O1 becoming the dominant pulsation mode instead of the FM \citep{Dupree2022,Jadlovsky2024}. With the incorporation of new observations, the average periods of both modes shifted slightly relative to the values found by \citet{Joyce2020}: the O1 became longer, whereas the FM became shorter, matching the period found by \citet{Kiss2006}.

Figure \ref{fig:Gaia_LSP} shows the phase offsets between Betelgeuse's lightcurve and RV curve in the context of similar stars measured with Gaia. We compare Betelgeuse to a limited sample of LPV-candidate Gaia stars from \citet{Trabucchi2023} for which we could estimate the phase offsets. We also show the pulsational phase lags for classical Cepheids for comparison, including high-precision data for 87 stars from the work of \citet{Szabo2007}, and lower precision data from Gaia DR3 for about 800 stars for which phase data are included in the database directly \citep{Ripepi2023}.

Under the adiabatic assumption, one would expect radial pulsation to have a $-\pi/2$ offset where maximum brightness corresponds to maximum compression, i.e, to zero pulsation velocity on the descent from RV maximum. In reality, the phase offset is closer to zero ($\Delta\phi\approx 0$) for pulsating stars, because ionization removes heat energy during the compression phase, thus introducing a phase lag \citep{Castor1968,Szabo2007}. For Betelgeuse, we find a relatively large phase offset for the FM, although its amplitude dropped considerably after the Great Dimming, which makes the fit more uncertain. The phase offset for the O1 mode, which is currently dominating the light variations in Betelgeuse, agrees with that of the shorter-period pulsating stars.  

To characterize the LSP beyond the specific context of Betelgeuse,
we calculated the phase offsets for the LSP stars published by \citet{Nicholls2009}, where OGLE-III \textit{I}-band photometry was available. We also calculated phase offsets for a subset of stars that were proposed to be LSP candidates by \citet{Trabucchi2023}, based on the recent Gaia FPR (Focused Product Release). Despite the fact that the Gaia data are generally sparse
for these stars, we managed to process nearly a third ($\approx$400 stars) of the full sample. The distributions of the two sets in Figure~\ref{fig:Gaia_LSP} agree well, with most stars saddling the $-\pi/2$ value. We also found a few targets with offsets close to $(-)\pi$. 

Turning now to the binarity model of the LSP, the configuration proposed by \citet{Soszynski2021} postulates that minimum light would occur when the companion passes in front of the red giant, corresponding to the point at which the RV variation of the primary goes through zero on the descending branch. This would occur at $\Delta\phi\approx \pi/2$ or $-3\pi/2$ offset. However, as Figure~\ref{fig:Gaia_LSP} shows, we observe the opposite not only in Betelgeuse, but in the majority of the LSP stars.

%
%
\section{Proposed LSP Hypotheses}  
\label{sec:hypotheses}
In light of the observations discussed above,
we now review and reevaluate proposed hypotheses for LSPs, discussing the validity of each as it applies to Betelgeuse.

\subsection{Hypothesis 1: LSP is actually the FM}
\label{subsec:notFM}
Recently, \citet{Saio2023} proposed that the $\approx$2100-day periodicity observed in Betelgeuse is in fact Betelgeuse's fundamental pressure mode (FM) rather than an LSP---a claim that is in tension with a number of other recent investigations (\citealt{Joyce2020, Dupree2022, Neuhauser2022,Granzer2022, MacLeod2023, MolnarResearchNote}; see also \citealt{Wheeler2023} for a recent review). However, classifying the $\approx$400-d periodicity as the first overtone (O1) and the $\approx$2100-d periodicity as the FM cleanly resolves the question of the origin of the LSP in Betelgeuse by asserting that there simply isn't one. This possibility is worth investigating.

\citet{Joyce2020} provided evidence that the 2100-day periodicity cannot be the fundamental mode on the basis that a periodicity of this length is not activated by the $\kappa$-mechanism 
in either linear oscillation calculations performed with GYRE
\citep{Townsend2013, Townsend2018} or 1D hydrodynamic calculations performed with MESA \citep{Paxton2011, Paxton2013, Paxton2015, Paxton2018, Paxton2019, Jermyn2023}. 
Likewise, \citet{MacLeod2023} demonstrate that since the Great Dimming, Betelgeuse's fundamental mode oscillation was perturbed into a combination of overtones. As in \citet{Joyce2020}, \citet{MacLeod2023} identify the $\approx$400-d periodicity as the FM by performing adiabatic and non-adiabatic calculations with GYRE, exploring a larger range of spherical degrees than in \citet{Joyce2020} (see their Table 1). 
Similarly, \citet{Jadlovsky2024} examine the behavior of shockwaves in the aftermath of the dimming, coming to the same conclusion as \citet{MacLeod2023} that the dominant pulsation mode had transitioned from the FM to the first overtone.

In general, semiregular variable stars can exchange pulsational energy between $p$-modes, so an exchange between the FM and the O1 is possible, but an exchange between the FM and LSP is not. In particular, energy exchange between modes happens in semiregular stars for period ratios close to a 2:1 resonant ratio, which is the ratio of the
$\approx$400-d and $\approx$200-d periods of Betelgeuse \citep{Buchler2004}.
The fact that there is mode switch from the $\approx$400-day periodicity to a $\approx$200-d periodicity being observed right after the Great Dimming (also \citealt{MacLeod2023}), rather than a mode switch at the $\approx$2100-day periodicity, is further evidence that the 400-d periodicity is the FM (as is the fact that 2100:400 is a $\approx$5:1 rather than 2:1 ratio).

Another critical issue with the 2100d-as-FM hypothesis is the stellar radius required to sustain an acoustic mode with a period of $\approx$2100 days. As
noted in \citet{Saio2023}, a radius of roughly $1300R_\odot$ is implied by models that treat Betelgeuse's $2100$-day periodicity as the FM. Observational constraints on Betelgeuse's radius provided by \citet{Haubois-2009}, \citet{Montarges-2014}, and \citet{KervellaDecin2018}
are in tension with these calculations, even at the longest distance measurements, as discussed in detail in \citet{MolnarResearchNote}. Measurements that find angular diameters larger than 50 mas (corresponding to radii $>1100R_\odot$) were done at longer wavelength and thus sample the MOLsphere (the close molecular layers) above the photosphere \citep{o'gorman-2017,Cannon-2023}. 

Still other studies undermine the 2100d-as-FM hypothesis on strictly evolutionary grounds. 
For example, in an investigation focused on constraining Betelgeuse's present-day evolutionary stage, \citet{Neuhauser2022} argue that Betelgeuse recently crossed the Hertzsprung gap, supported by its historically documented color change over the past several thousand years. A recent traversing of the Hertzsprung gap places Betelgeuse in the early stages of core helium burning, which is incompatible with the core-carbon-burning evolutionary phase implied by a 2100-day fundamental mode \citep{Saio2023}. 

Given the range of arguments against a 2100-d fundamental mode period from many orthogonal perspectives, we reject the claim that the pulsation mode widely classified as Betelgeuse's LSP is actually the star's FM. We proceed henceforth referring to the $\approx400$-d periodicity as the FM and the $\approx2000$-d periodicity as the LSP in Betelgeuse.  While the case of Betelgeuse alone is enough to debunk the 2100d-as-FM hypothesis as a universal explanation for LSPs, the existence of Sequence D on the PL diagram (see Figure \ref{fig:PL}) does so more conclusively.

\subsection{Hypothesis 2: Giant convective cells}
\label{subsec:convection}

In light of the evidence that LSPs are not simply mis-identified FMs, as discussed above, \citet{Stothers1971} proposed the hypothesis that giant convective cells can explain the LSP phenomenon, generalized by \citet{Stothers2010} to extend to all LSP pulsators. Broadly, while the acoustic FM timescale is limited by $\sim{}R/c_\mathrm{s}$ and thus predicts periods too short to explain the LSP for reasonable stellar radii and $\Teff$, a hypothetical periodic signal might be mediated by the overturn timescale of giant convective cells, 
\begin{equation}
t_\mathrm{conv}=\ell/\vc
\end{equation}
where $\ell$ is the plume size and $\vc$ is the characteristic velocity of convection. In the limit of low $\vc<c_s$ with a maximum plume size $\ell$ approaching $R$, this timescale can indeed be longer than the FM.

This picture has order-of-magnitude agreement in the luminosities and radial velocity amplitudes \citep{Stothers2010}. Following Mixing Length Theory (MLT, \citealt{BohmVitense1958}; see recent review by \citealt{Joyce2023}), the local characteristic length scale of convection can be expressed as $\ell=\alpha_\mathrm{MLT}H_P$ where $H_P$ is the pressure scale height and $\alpha_\mathrm{MLT}\approx2-4$ for RSGs (e.g. \citealt{Chun2018, Goldberg2022a}).  
In MLT, the $\vc$ can then be estimated \citep{Kippenhahn} from the superadiabaticity $\grad-\grade$, the local gravity $g$, and thermodynamic properties
 \begin{equation}
     \vc^2=gQ\left(\grad-\grade\right)\frac{\ell^2}{\nu H}.
     \label{eq:mltv}
 \end{equation}
 where $\nu$ is a geometric factor encoding plume geometry (typically $\nu=8$ following \citealt{Henyey1965} and others), $Q=-\mathrm{D}\ln{}T/\mathrm{D}\ln\rho$ is determined by the equation of state (EOS), and $\grad-\grade$ can be related to the flux carried by convection ($F_\mathrm{conv}$) and the mixing length by
 \begin{equation}
     F_\mathrm{conv}=\rho c_{P}T \sqrt{gQ}\frac{\ell^2}{\sqrt{\nu}}H^{-3/2}(\nabla-\nabla_\mathrm{e})^{3/2},
     \label{eq:mltflux}
 \end{equation}
and where $c_P$ is the specific heat at constant pressure.

In the extended convective envelopes of RGB stars, AGB stars, and RSGs, indeed $\ell\sim{}R$. In such a star, large $H_P/r\sim0.25$ in the interior yields large-scale plumes spanning a sizeable fraction of the stellar radius, as recovered both by simulations \citep[e.g.][]{Freytag2002,Dorch2004,Brun2009,Plez2013,Chiavassa2009,Chiavassa2010,Chiavassa2010b,Chiavassa2011,Chiavassa2011b,Chiavassa2017,Chiavassa2024,Goldberg2022a,Antoni2022} and inferred from observations \citep[e.g.][]{Montarges2017,Chiavassa2018,Chiavassa2018b,Kravchenko2018,Kravchenko2019,Kravchenko2020,Kravchenko2021,Zhang2024}. 
For RSGs like Betelgeuse, MLT predicts convective velocities of $\sim5-25$ km/s for RSG stars 
\citep[see discussions in][]{Joyce2020,Goldberg2022a}, somewhat consistent with the required convective velocity to give rise to an overturn time of period $P$ in a star of radius $R$
\begin{equation}
v_{c,\mathrm{LSP}}\equiv{}\frac{R}{P}=2.8\ \mathrm{km/s}\left(\frac{R}{764\Rsun}\right)\left(\frac{P}{2170\mathrm{d}}\right)^{-1}.
\end{equation}

However, while there is order-of-magnitude agreement with the observed LSP, tensions arise.
Large-scale convection is more or less ubiquitous in cool, luminous giants, and as such, we would expect all such stars to exhibit similar LSP behavior if giant convective cells were the cause. In reality, only 30\% do \citep{Nicholls2009, Soszynski2021}.
Moreover, convective motions do not remain coherent forever, even in the case of a giant convective cell, whereas LSPs are generally steady in nature. \citet{Takayama2023}, for example, found that of $\approx9000$ LSP candidates in the LMC and SMC, only $\approx150$ of those candidates showed any evidence for a changing LSP duration over the observing term. In the convective picture, over time significant stochasticity is expected to be introduced into both the lightcurves and observed velocities, and eventually the stochastic variability should wipe out a periodic signal from convection alone. In fact, in the hotter OB star regime, variability from subsurface convection has been proposed to better explain stochastic low-frequency variability \citep{Schultz2020,Cantiello2021}, and in RSGs, similar stochastic variability from granulation has been recovered from LMC observations \citep[e.g.][]{Zhang2024}.

Giant convective cell models also make no prediction for RV modulation on the timescale of LSP.
The convective velocity $v_c$ is a characteristic velocity of convective transport and \textit{not} simply a bulk fluid velocity, as plumes move upwards and downwards and laterally at the surface \citep{Stein1989,Stein1998}. In fact, the convective overturn time is often considered to be a \textit{decoherence} time. In 3D simulations of RSG envelopes \citet{Goldberg2022a} found decoherence times on the order of $\approx$300-500 days. Additionally, \citet{Ma2024} show that convective plumes in an RSG-like convective envelopes could mimic a 5 km/s signal inferred to be surface rotation \citep[see, e.g.][]{Kervella2018}. However, they find that the coherence time and dipole orientation of such a signal are significantly less than their 5 years of simulation time, with snapshots showing synthetic $v$sin$i$ dipoles ranging from $<1$km/s to 10km/s at random orientations. 

The MLT treatment which defines a single `convective velocity' scale and thus a single `convective overturn' timescale is, of course, a dramatic simplification of the true complexities of convection (see e.g. \citealt{Joyce2023} \S5, as well as the discussion in \S5 of \citealt{Goldberg2022a} in the case of RSGs, especially their Fig.~19). 
Stars like Betelgeuse with large envelopes hosting vigorous convection are indeed more poorly approximated by MLT than solar-like stars by virtue of their significant deviation from solar conditions (see e.g. \citealt{Joyce2018}). Locally Super-Eddington luminosities due to the presence of strong opacity peaks lead to numerical as well as physical challenges for 1D models, which must be overcome through various ``engineering" techniques (see, e.g., \S7.7 of \citealt{Paxton2013} and \S7.2 of \citealt{Jermyn2023}; and discussions and references therein). 
Moreover, the convective motions observed in 3D Radiation-Hydrodynamics simulations of near-surface convection in RSG envelopes other luminous massive stars do not even resemble motions typical of MLT-like convection out to the surface \citep{Schultz2020, Schultz2022, Goldberg2022a}; above some radius ($R_\mathrm{corr}$) the correlations between the radial fluid velocity and the density, entropy, and opacity invert -- cold, dense, high-opacity material which would sink in MLT-like convection is correlated with \textit{positive} $v_r$ -- even at high optical depth $\tau \gtapprox 100$.\footnote{
While there are significant caveats with MLT in the context of defining a single persistent convective overturn time, it is unlikely that the simplifying assumptions of MLT discussed above would affect results in this work in any other way. As noted by \citet{Joyce2020}, the primary impact of changing $\alpha_\mathrm{MLT}$ in models of 10-25$M_\odot$ was to shift in temperature the location of the red giant branch, leading to a factor-of-ten increase in the uncertainty of the effective temperature constraint as compared to the error bars reported observationally by \citet{Levesque2020}. 
While uncertainty in the optimal value of the mixing length could certainly impact results through $\alpha_\mathrm{MLT}$’s impact on the synthetic seismic frequency spectra against which Betelgeuse’s pulsation spectrum was compared, this uncertainty was fully accounted for in \citet{Joyce2020}, whose parameters we use elsewhere in our analysis.}


At the very least, any modulated RV signal caused by giant convective cell turnover is likely to be far more complex than the RV modulation seen in Betelgeuse or the Gaia LSP variables, and an LSP driven by large-scale turbulent motions would not persist for the $\gtrsim$hundred years over which Betelgeuse has been observed.
 
\subsection{Hypothesis 3: Mode-interactions (beats)}
\label{subsec:beats}
Interaction between close-by modes can create long beating periods that could match the LSPs. However, beating modulates the pulsation amplitudes and hence the lightcurve envelopes, not in the average brightness. Furthermore, if the LSP period were a beating period between modes in Betelgeuse, then it would correspond to a physical signal close to the pulsation frequencies. This signal would be offset by about 19\% and 10\% relative to the FM or the O1 frequencies, instead being of an independent signal at $1/P_{\rm LSP}$.

Another aspect of mode interaction concerns interactions between the LSP and the other pulsation modes. If two (or more) strong pulsation modes are present in the star, they will both influence the envelope structure in which the other modes have to travel. These interactions would cause combination frequencies to appear in the frequency spectrum of the star, as seen in double-mode Cepheids and RR Lyrae stars \citep[see, e.g.][]{gruberbauer-2007}. Such combination frequencies with the LSP would be very similar to the beating frequencies mentioned above (e.g., at $f_{\rm FM}\pm f_{\rm LSP}$). But such signals have not been reported for Betelgeuse, which suggests that the LSP is not the product of a beating phenomenon.

\subsection{Hypothesis 4: Rotation}
\label{subsec:rotation}
As noted in \citet{Wood2004IAU}, if stars exhibiting LSPs are ``rotating prolate spheroids,'' meaning they are distorted in such a way that the polar radius is larger than the equatorial radius, they would produce the velocity curves characteristic of LSPs. However, the rotational velocities that would be required for this explanation are far too high compared to observations: the apparent 5-15 km/s surface velocity of Betelgeuse corresponds to a 30-year periodicity, whereas Betelgeuse's LSP has a 5.5 year periodicity. Further, LSPs fall on a period-luminosity sequence, and it is not clear why this would happen if (solid body) rotation were the explanation.

\subsubsection{Differential Rotation}
On the main sequence, the cores of high-mass stars have been predicted by some models to rotate about 100x as fast as their surfaces (see, e.g., 
\citealt{Heger2000,Heger2005}), though more recent studies have shown that these models over-predicted core rotation by a factor of 10. In one such study, \citet{Pedersen2022} found that measurements indicate that SPB stars with masses of 5-10 $M_\odot$ rotate close to rigidly. Conversely, \citet{Hermes2017} do find evidence of a connection between high stellar birth mass and rapid core rotation via asteroseismology of white dwarfs, although again for masses below that of Betelgeuse.
There are currently no measurements of core rotation in stars above $10\,M_\odot$, so extrapolation to Betelgeuse should be taken with caution.

Further, the relationship between envelope and core rotation along the main sequence is not necessarily preserved once the star enters the core helium-burning phase. 
\citet{Mosser2024} find that, for low-mass red giants, the evolution of the core rotation rate in core-helium-burning stars scales with the inverse square of the stellar radius, in good agreement with \citet{Tayar2019}.
Although Betelgeuse is a massive star, and the core rotation rates of red supergiants have not been probed directly, asteroseismic relations for low-mass, core helium-burning red giants are informative for this case.
The model described in \citet{Mosser2024} is that of a rigid, dense helium core surrounded by a radiative layer with smooth rotation profile, all encased by a rigid convective envelope: a picture that describes Betelgeuse reasonably well. Putting the \citet{Joyce2020} lower mass bound ($\sim16.5 M_\odot$) and radius ($764 R_\odot$) for Betelgeuse into \citet{Mosser2024}'s equation 3 yields a core rotation rate of approximately 400 days. This is much shorter than the LSP, so modulation by internal rotational is ruled out.

Though \citet{Lovekin2009} note that extreme cases of differential rotation will cause slight shifts in acoustic mode frequencies, the extent of such a shift would be nowhere near the factor of $\approx5$ decrease in frequency required to make the FM mimic the LSP.
Further, \citet{JermynTayar2020} find that there is very little radial differential rotation in stars with convective envelopes (in both short and long period eclipsing binaries) anyway.

\subsection{Hypothesis 5: Magnetism}
\label{subsec:magnetism}
Various authors have explored the possibility that a process analogous to the magnetic solar cycle---e.g. star spots, convectively-driven magnetic dynamo---could generate LSPs (e.g. \citealt{Soker1999} in AGBs and \citealt{Wood2004} in RSGs), but several problems arise with this theory \citep{Olivier2003}. 
Of particular importance is the fact that spots on a rotating surface are not capable of explaining lightcurve variations across the appropriate range of colors (i.e. the observed chromatic behavior, or chromaticity), as discussed in \citet{Wood2004} and more recently in \citet{Takayama2015}.

Another issue is that magnetic cycles in stars like Betelgeuse are expected to be much longer than 11 years.
Although there is no simple means of scaling the solar magnetic cycle to red supergiants, and dynamo cycles have not been observationally confirmed in stars like Betelgeuse, we know that the period of the magnetic solar cycle goes as
\begin{equation}
    T_\mathrm{solar} \sim \frac{R^2}{\eta},
\end{equation}
where $R$ is the stellar radius and $\eta$ is the magnetic diffusivity. The magnetic diffusivity is a function of the scale of the convective motions in the envelope and the star's conductivity. Every term in this relation contributes to increasing the period of the magnetic cycle in Betelgeuse relative to that of the Sun.  
Given that the period of the LSP is roughly $5.5 < 11$ years, a magnetic cycle similar to the Sun's is not a viable explanation for the LSP in a large, evolved star with a deep convective envelope.

Some magnetic phenomena affecting high-mass stars are worth considering.
The interplay between differential rotation and magnetic fields has been studied primarily in OB stars \citep{MaederMeynet2011, MaederMeynet2012}, which was Betelgeuse's spectral type when it was on the main sequence.
\citep{Keszthelyi2019} find that fossil fields in massive stars can induce mass-loss quenching and magnetic braking; however, episodic mass loss is periodic and should happen at the frequency of the FM. The role of fossil fields would be to suppress the amount of mass dispersed per episode, not to alter the frequency of mass-loss episodes. 
Fossil fields therefore do not cause the LSP.
Finally, while shorter-period, dynamo-generated magnetic fields can exist in the sub-surface convection zones in hot, massive stars \citep{Cantiello2011}, Betelgeuse is not hot enough to have an iron opacity peak near the surface.

\subsection{Hypothesis 6: Non-radial Pulsations}
\label{subsect:nonradpulse}
Any pulsation-related explanation for the LSP must involve a driving mechanism that produces pulsations below the frequency of the FM (see Section \ref{subsec:notFM}), which means it cannot be acoustically driven. Such drivers may be intrinsic or extrinsic to the star. Here we address the possibility that the LSP is caused by an \textit{intrinsic} source of non-radial pulsation.

\subsubsection{Strange Modes}
Strange modes are unstable modes with potentially high growth rates that occur in extremely non-adiabatic environments dominated by radiative energy transport \citep{Saio2009}. They are caused by a large difference in phase between the density and pressure maximum in the oscillating partial ionization zone.
Strange modes can be found in the radial spectra of many luminous stars, including Cepheids, RR Lyrae, LBVs, and other high-amplitude pulsators \citep{Buchler1997}. Their presence in luminous red giants is discussed in detail in \citet{WoodOlivier2014}. 

Strange modes with LSP-like periods have been found in red giant models, arising due to the interaction of stellar oscillations with convective energy transport \citep{Wood2000}. However, such modes are highly damped, and models predict an exponential decay timescale of less than 100 days upon perturbation \citep{Wood2004}.
They are therefore highly unlikely to be observed in real stars. 

More recently, \citet{Saio2015} argued that such oscillatory convective modes can, in fact, explain LSPs in real red giants. However, to generate modes of the correct period to reproduce Sequence D (see Figure \ref{fig:PL}), \citet{Saio2015} engineered highly non-adiabatic stellar models dominated by radiative transport in their outermost layers, which do not accurately reflect the physics of red supergiants. Three-dimensional simulations show that in fully convective RSG envelopes, the convective velocities penetrate above the surface. Large turbulent motions well outside stellar photosphere form coherent plumes, meaning that while radiation carries much of the flux, the convective motions persist far above the photosphere \citep[see, e.g.][]{Chiavassa2009, Chiavassa2011b, Chiavassa2024, Goldberg2022a, Ma2024}. In fact, at any radius defined as the ``photosphere,'' there is even some fluid that is optically thick and retaining its entropy such that its behavior is convective even though radiation is carrying flux elsewhere along in the near-surface layers (see discussion in Section~4.4 of \citealt{Goldberg2022a}). 

\citet{Saio2015} encoded the surface-radiation-dominated physics in their one-dimensional models (computed with MESA) by using a low, very sub-solar value for the convective efficiency ($\alpha_\text{MLT}/\alpha_{\text{MLT},\odot} = 1.2/1.8 = 0.67$) in the MLT formalism, and it is this choice that enables both the large surface radiative zones and large radii of their models. 
In reality, convection is more efficient in a red supergiant's envelope than it is in the Sun's \citep[e.g.][]{Chun2018}, with typical values of $\alpha_\text{MLT}\approx2-3$ inferred from observed HR positions of nearby RSGs. Further, use of $\alpha_\text{MLT} = 1.2$ produces models with inappropriately low effective temperatures \citep{Levesque2020}.

Given the inability of models to reproduce simultaneously the correct RSG temperatures, convective envelope physics consistent with RSGs, LSP-appropriate strange mode frequencies, and strange mode lifetimes long enough to be observable, we reject strange modes as the explanation for the LSP of Betelgeuse. 

\subsubsection{g-modes }
Because they can have longer periods than the radial fundamental mode, $g$-mode pulsations were first proposed as an explanation for the LSP by \citet{Wood1999}. However, the structure of red supergiant envelopes once again undermines the viability of this explanation.
As noted by \citet{Takayama2015} and discussed earlier in \citet{Wood2004}, $g$-modes must develop in a radiative zone. At best, LSP stars have a very thin outer radiative layer above their large convective envelopes. The size of this region strongly limits the amplitude of $g$-modes that could develop there, and any such modes would be too small to explain the large amplitudes of both the brightness and radial velocity variations that characterize LSPs.  

Inverting the argument, any $g$-mode that would be observationally detectable as the LSP in Betelgeuse implies the existence of a
quiet, ``well-behaved,'' and large-enough radiative zone near the surface of Betelgeuse in which this mode is propagating. Neither stellar evolution calculations nor models of large convective cells in red supergiants suggest this is possible. 

To comment briefly upon $g$-modes in the stellar core, we note that $g$-modes are evanescent (i.e. exhibiting amplitude decay over distance) throughout the convective envelope--which is, once again, enormous in Betelgeuse--and would therefore not be detectable at the stellar surface. We are thus left with the fact that $g$-modes can account for LSP periods but not LSP amplitudes, no matter their origin in RSGs. 

\subsection{Hypothesis 7: $\kappa$-mechanism in the dust shell}
\label{subsect:kappadust} 
The final remaining pulsational scenario consistent with the LSP timescale posits a radial pulsation from a mechanism driven outside the optical stellar photosphere, dragging both the photosphere and the external material. \citet{Winters1994} and \citet{Hoefner1995} argue that opacity changes in the dust shell surrounding the star could cause oscillations via a ``dust-$\kappa$" mechanism, whereby the strong temperature and density sensitivity of the sublimation state (and thus opacity) of circumstellar dust could induce radial pulsations in the dust shell(s) \citep[see also discussions by][]{Wood2004}.

From a period-mean-density relation with generic period $\Pi$, 
\begin{equation}
\Pi \propto \frac{1}{\sqrt{G\bar{\rho}}} \propto R^{3/2}
\end{equation}
where $\bar{\rho}$ is the average density and the second proportionality assumes the mass scale is dominated by the stellar material and not the dust.\footnote{In order to achieve a modulated RV signal, near stellar-photosphere material would need to take part in the oscillation, so the stellar mass is expected to set the mean density unless there is very appreciable mass within the dust shell.} For an LSP:FM period ratio of $\approx5$, assuming the FM is an oscillation at the (optical or IR) stellar photosphere, this would mean that the `outer' radius mediating the RV signal would be $\approx3$ times the stellar radius, or $\approx2200R_\odot$ for a radius of $764R_\odot$ \citep{Joyce2020}. This is consistent with placement inside the region probed by radio observations indicating the presence of significant circumstellar dust from $2-6$ stellar radii \citep{Altenhoff1979,Lim1998,Harper2001,Smith2009,OGorman2015}.

A dust-$\kappa$-like mechanism is also appealing because it simultaneously predicts the chromatic behavior of the LSPs and could explain IR lightcurve dips apparently modulated by increases and decreases in dust activity in front of the star \citep{Soszynski2007,Soszynski2021,Percy2023}. Such a pulsation, if it is indeed dragging the whole star, would also explain the typical LSP LC-RV phase offset, with the LSP luminosity minimum (indicating more dust) occuring shortly after the maximum in the modulated stellar RV (indicated by low-but-increasing photospheric RV).

Ultimately, the ability of such a pulsation out at 3$R$ to adequately modulate the RV at the level of $\approx\pm2.5$km/s at the stellar surface is the primary point of tension for the dust-$\kappa$ theory. 
As pointed out by \citet{Wood2004IAU,Wood2004}, if dust is being created and ejected via a dust-$\kappa$ mechanism, this phenomena would occur primarily outside the stellar photosphere and it is unclear how the deeper photospheric layers could be dragged enough to cause the RV amplitude modulations seen in LSPs, and this argument extends also to Betelgeuse. Moreover, changes in $L$ and the optical stellar photosphere $R$ would be accompanied by large changes in $T_\mathrm{eff}$. We note that this was not seen even in the Great Dimming's extreme dust ejection near LSP minimum \citep{Montarges2021}, during which Betelgeuse's $T_\mathrm{eff}$ cooled slightly but remained at $\Teff\approx3600\pm25$K \citep{Levesque2020}. This low variation in $\Teff$ is consistent with other LSP variables, which show little modulation in $\Teff$ \citep{Wood2004}. Finally, though Betelgeuse does exist within 
a surrounding dusty CSM, it should be noted that some LSP stars show little evidence for large circumstellar dust shells to begin with \citep{Hinkle2002,Olivier2003}. 
While some kind of externally-sourced pulsation remains capable in theory of reconciling the RV phase and the LSP, there is no clear way to do this
without large corresponding changes in $\Teff$.

\subsection{Hypothesis 8: Modulation by a companion }
\label{subsec:binary}

In this section, we overview the ways in which a companion could induce an LSP \citep[see, e.g. ][ and discussions and references therein]{Wood1999,Wood2004,Retter2005AAS,Soszynski2007,Soszynski2021, Percy2023}. This family of solutions is the generally favored mechanism for producing LSPs, as this class of scenarios naturally explains the long timescale and steady period as the orbital time of the companion. As discussed in Section~\ref{sec:LSPs}, the existence of a P-L relation for LSPs is not an intrinsic problem, as orbital architecture is sensitive to the stellar mass and radius. A remaining point of tension is the relatively high fraction of long-period variables with LSPs and RV modulations compared to main-sequence stars \citep{Nicholls2009}, which can be addressed by appealing to mass growth of the orbiting body (e.g. \citealt{Livio1984}; see discussion in \S5 of \citealt{Soszynski2021}). At least regarding 
RSGs, the observed binary fraction is very high for massive stars \citep{Sana2012,deMink2013}, and this fraction does not include undetected lower-mass companions (we return to this in Section~\ref{subsec:observability}).

\subsubsection{Hypothesis 8a: Tidally-Induced Oscillations  } 
\label{subsec:tides}
Tidally-induced oscillations are another form of non-radial oscillation (see Section \ref{subsect:nonradpulse}) that emerge in highly non-adiabatic conditions (see, e.g., \citealt{Bunting2019, Sun2023}). 
In the case of a companion inducing modes on Betelgeuse via tidal forces, it is the companion's orbital timescale that could modulate the RV variations, but an oscillation of the star or system that determines the luminosity fluctuations.
If we consider tidally-driven modes induced on Betelgeuse by a companion, these modes are still acoustic $p$-modes acting over the surface of Betelgeuse and are therefore limited by its radius ($R/c_s$). Considering the case of a severely tidally distorted Betelgeuse, because $R/c_s=t_\mathrm{dyn}$ at the surface,  to distort the star substantially in a Roche potential would require a very short orbital period on the order of the FM. As such, tidally-driven modes in the star itself cannot have significantly lower frequencies than the fundamental mode.

If, on the other hand, we consider a scenario in which the potential of the orbital system---Betelgeuse and its companion---drives a pulsational mode that acts over the radius of the whole system (rather than Betelgeuse alone), it is unclear what would be pulsating. If the oscillating medium is a steady-state dust shell encompassing both bodies, then we encounter difficulties providing a physical justification for a pulsation in such a low-density environment (see also Section~\ref{subsect:kappadust}).
There remains the possibility that such an oscillation might mediate the properties of the dust, thereby modulating the luminosity and chromatic behavior. If the orbit provides RV modulation in 1:1 resonance with an induced oscillation period, this might explain why there could be a sizeable RV modulation with relatively steady $\Teff$. In general, however, this scenario would not prefer any particular phase relationship between the LSP and the RV, since a pulsating dust shell modulating the luminosity would not know about the observer's direction. 
In either case, tidally-induced oscillations do not provide a clean explanation for the LSP.

\subsubsection{Hypothesis 8b: Occultation } \label{subsec:occultation}
The current preferred hypothesis for LSPs posits that a low-mass companion is dragging a cloud of dust and gas along the observer's line of sight, temporarily blocking the view of the star and causing the luminosity minimum \citep[see, e.g.][]{Wood1999IAU,Wood2004,Soszynski2007,Soszynski2014,Soszynski2021,Percy2023}. 
This theory is appealing as it 
\\(1) naturally explains a timescale longer than the FM, 
\\(2) allows for RV modulation on the LSP timescale 
\\(3) without substantial changes in $\Teff$, and 
\\(4) explains dust-like chromatic behavior. 
As discussed in detail by \citet{Soszynski2021}, a compelling piece of evidence for dust modulation on an orbital timescale as an explanation for the LSP phenomenon comes from inspecting the shape of the LSP lightcurves. These vary in the characteristics of their luminosity fluctuations, from narrower dips commensurate with small orbiting dust blobs, through deep periodic signals consistent with much larger or more complex orbiting dusty bodies. As discussed in Section~\ref{sec:LSPs}, there is no significant tension between a companion hypothesis and the existence of a P-L relation; in fact, this succinctly explains why Sequence D in our Figure~\ref{fig:PL} and Fig.~2 of \citet{Soszynski2021} extends from the sequence of lower-luminosity eclipsing binaries (sequence E). 

A diversity of LSP amplitudes would be expected for different inclination angles and companion/dust cloud properties, and this model can explain cases where there is no detectable RV signal by invoking a very high mass ratio. However, the occultation hypothesis predicts that when RV modulation is present on the LSP timescale, one should infer the orbiting companion to be somewhere along the observer's line of sight, i.e., the star's $v_r$ should be near-average and accelerating \textit{towards} the observer near luminosity minimum, at a phase offset of $\phi\approx+\pi/2$ or $-3\pi/2$ (see Figure~\ref{fig:Gaia_LSP}).

Unfortunately, this prediction does not hold for Betelgeuse with its LSP-RV phase offset of $-0.63\pi$. This is almost $180^\circ$ out of phase with what would be expected for dust gravitationally bound to an orbiting body or at the L1 Lagrange point and still $\approx120^\circ$ out of phase with dust sitting the L4 or L5 Lagrange points. The occultation hypothesis is also in tension with the RV phases for the majority of the Gaia sample shown in Figure~\ref{fig:Gaia_LSP} and discussed in Section~\ref{sec:phaseoffset}.
Thus while the orbiting dusty body hypothesis can explain the RV amplitude and LC period in Betelgeuse, it does not explain their phase offset.

\subsubsection{Hypothesis 8c: Circumstellar Modulation Near Companion }
\label{sec:buddyscenario} 
\bgroup\def\lowerthis#1 {\smash{\lower6pt\hbox{#1}}}
\begin{table*}[]
\centering
\begin{tabular}{lccccccc}
\hline
\lowerthis Hypothesis           & \lowerthis Section & \lowerthis Timescale & Properties & Low variation & Dust-like & LC-RV & \lowerthis Persistence \\[-3pt]
& & &  of RV & in $T_\mathrm{eff}$  & chromaticity &  offset & \\\hline
Misidentified FM               & \ref{subsec:notFM}        & \checkmark+                             & X                                    & ?                                          & -                                           & ?                                 & \checkmark+                                \\ 
Giant convective cells          & \ref{subsec:convection}        & \checkmark                              & ?                                    & \checkmark                                         & -                                           & -                                 & X                                \\ 
Mode interactions    & \ref{subsec:beats}        & X                              & X                                    & -                                          & -                                           & -                                 & X                                \\ 
Rotation             & \ref{subsec:rotation}        & X                            & \checkmark                                   & ?                                          & ?                                           & \checkmark                                & \checkmark+                               \\ 
Magnetism            & \ref{subsec:magnetism}        & X                              & -                                    & X                                          & X                                           & ?                                 & ?                                \\ 
Non-radial pulsation & \ref{subsect:nonradpulse}        & \checkmark+                             & \checkmark                                     & ?                                          & ?                                           & ?                                 & X                                \\ 
Dust-$\kappa$           & \ref{subsect:kappadust}        & \checkmark+                             & \checkmark+                                   & X                                          & \checkmark+                                          & \checkmark+                                & \checkmark                                \\ 
Binarity: tidal                & \ref{subsec:tides}        & X                             & \checkmark+                                   & ?                                          & ?                                           & ?                                 & \checkmark+                               \\ 
Binarity: occultation          & \ref{subsec:occultation}        & \checkmark+                             & \checkmark+                                   & \checkmark+                                         & \checkmark+                                          & X                                 & \checkmark+                               \\ 
Binarity: dust modulation     & \ref{sec:buddyscenario}        & \checkmark+                             & \checkmark+                                   & \checkmark+                                         & \checkmark+                                          & \checkmark+                                & \checkmark+                               \\ \hline
\end{tabular}
\caption{Summary of hypotheses for Betelgeuse's LSP.
Physical properties and observational constraints/pieces of information we have for Betelgeuse are listed across the top row.
A checkmark-plus sign pair (\checkmark$+$) underneath one of these pieces of information
 indicates that the corresponding hypothesis, given in the left-most column, is supported by that piece of information. A checkmark (\checkmark) indicates consistency. A question mark (?) indicates that the relationship between the hypothesis and constraint is unclear or unknown. A dash (-) means that the hypothesis does not provide any information on the constraint. An X indicates that the hypothesis is contradicted by the constraint. }
\end{table*}
\egroup
\label{table:summary}

The tension between the dusty companion hypothesis and the RV/lightcurve phase offset requires reevaluating the broader physical picture. From Figure~\ref{fig:lightcurve_vs_RV}, the phase difference between the lightcurve and RV modulation favors a scenario in which the photosphere begins to move away from the observer near the time of LSP luminosity minimum. For this to be caused by a companion, the companion must be 180 degrees out of phase with the dust that induces the brightness minimum of the LSP. Thus we propose a new scenario: a companion whose orbital period sets the timescale of the LSP and RV fluctuations, and whose presence modulates, removes, eliminates, or changes the optical depth of any potentially dusty material in its vicinity, rather than dragging a trailing cloud of dusty material. There are multiple ways in which this might occur, including, for example, dynamically perturbing the potential of the dusty medium or through irradiation.

For this scenario to be feasible, there must be material for the companion to interact with,
and the timescale over which the circumstellar matter replenishes and enters into proximity with the companion must be shorter than the orbital period. Both of these conditions appear to be satisfied in Betelgeuse. As supported by observational measurements of material extending from the MOLsphere to many tens of stellar radii beyond, it is well-established that dusty material exists outside the photosphere \citep{Smith2009,Kervella2009,Kervella2011,Montarges-2014,OGorman2015,Dharmawardena2018,Haubois2019}.

The connection between binarity, convection, pulsation, dust condensation, and mass loss has been explored by various groups in both simulations and observations (see, e.g., \citealt{Hofner2018} for a recent review in the context of AGB mass loss). 
Both 1D and 3D simulations of evolved cool giant stars exhibit dust-driven winds forming at a few times the stellar radius, seeded by (asymmetric) material which gets shocked and levitated by the pulsating stellar surface (see, e.g., \citealt{Freytag2008,Freytag2017,Freytag2023,Liljegren2016,Liljegren2017,Liljegren2018,Bladh2019a,Bladh2019,Siess2022,Steinwandel2022,Wiegert2024}). The outflows in these works exhibit a wide range of circumstellar densities and mass-loss rates depending properties near the stellar surface, such as the pressure scale height, temperature, pulsation period, and degree of asymmetry (as well as composition; see, e.g. \citealt{Karakas2014, Lattanzio2016}).

Likewise, observations indicate a correspondence between the onset of dusty winds and different pulsation sequences \citep[e.g.][]{McDonald2019, Yu2021}. LSPs have long been associated with dusty activity, as \citet{WoodNicholls2009} found increased mid-IR emission with no difference in color for RGB stars with LSPs compared to those without, interpreted to indicate the presence of dust in a clumpy or disk-like configuration.
It has also been shown in hydrodynamic simulations of AGB star winds that a companion may modify the circumstellar matter formed by dusty outflows \citep{Maes2021}, with differing geometry depending on the properties of the outflow and companion.

Two important timescales may set the launching of periodic or stochastic outflows which might turn into dusty fodder: pulsations and convection. 
The pulsation timescale (FM or O1) is necessarily shorter than the LSP, and the coupling of near-sonic, large-scale convection in a pulsating envelope has been hypothesized to drive stellar winds \citet{Yoon2010}. Similarly, the convective motions themselves may lead to steepening shocks in the atmosphere. This may stratify material beyond the stellar photosphere, well past the dust formation radius, with variations changing over the convective turnover time \citet{Fuller2024}. 
A timescale for this process can be estimated by considering motions in the deep interior, following equation B4 of \citet{Ma2024},
\begin{equation}
t_{\mathrm{turnover}} \sim \frac{R}{v_c}\sim 0.8\mathrm{yr}\left(\frac{R}{800 R_{\odot}}\right)\left(\frac{T_{\text {eff }}}{3600\mathrm{~K}}\right)^{-\frac{4}{3}}\rho_{s,-9}^{\frac{1}{3}},
\end{equation}
where $\rho_{s,-9}$ is the density at the optical photosphere $\rho/10^{-9}$ g/cm$^3$, or surface timescales much shorter than that (See discussion in Appendix B of \citealt{Ma2024}).
It is also known that for particularly extreme dust ejection events, such as the Great Dimming \citep{Montarges2021,Dupree2022,Drevon2024}
a major dust cloud forms and passes on timescales shorter than the LSP \citep[see extensive discussion in][]{MacLeod2023}. 

The modification we are presenting to the dust-modulating-companion scenario originally discussed in, e.g., \citet{Wood1999,Soszynski2007,Soszynski2014,Soszynski2021} is appealing for the four reasons mentioned in Section~\ref{subsec:occultation} (timescale, RV modulation, low variance in $\Teff$, and chromaticity),
with the additional benefit that it explains the phase offset in Betelgeuse, as well as many of the \textit{Gaia} LSP variables (Figure~\ref{fig:Gaia_LSP}). 

A dust-modulating companion also leaves open the possibility of secondary eclipses to be seen in the IR, which represent strong evidence for the binarity scenario in other cases \citep{Soszynski2021}. The primary difference is that the companion would be $\approx$180 degrees out of phase with the bulk of the circumstellar dust rather than dragging the dust along as in the occultation scenario. This phase difference would potentially but not necessarily yield differences in the  (highly-uncertain) expected dust geometry, depending in part on whether the dust modulation is dynamical or radiative in nature. 

We note that elimination or dispersal of dust near an orbiting body need not be the explanation for \textit{all} LSP variables. Many LSP variables do show steady amplitudes and periodic narrow lightcurve dips \citep{Soszynski2021}, and at least a few show RV modulation phases consistent with luminosity drops when the companion would be in front of the star. It is likely that the mass, luminosity, and $\Teff$ of the companion, as well as the replenishment timescale of material provided by the star, may determine whether the system drags and condenses dust along its orbit or creates an under-density in its surroundings. We leave it to future work to determine the exact conditions under which each orbitally-modulated LSP/RV phase scenario (occultation, dynamical dispersal, accretion, irradiation, etc.) could arise.

We thus conclude that circumstellar dust modulation by a companion is the most plausible explanation for the LSP in Betelgeuse.
Within this framework, we can use the observational data available to estimate the properties of a companion consistent with the RV and RV-LC phase difference. This companion is hereafter referred to as $\bbud$\footnote{``B" stands for Betelbuddy}. 
The properties (Section~\ref{sec:properties}) and detectability (Section~\ref{subsec:observability}) of $\bbud$ are discussed subsequently.

\begin{figure*}
\centering
\includegraphics[width=0.85\textwidth]{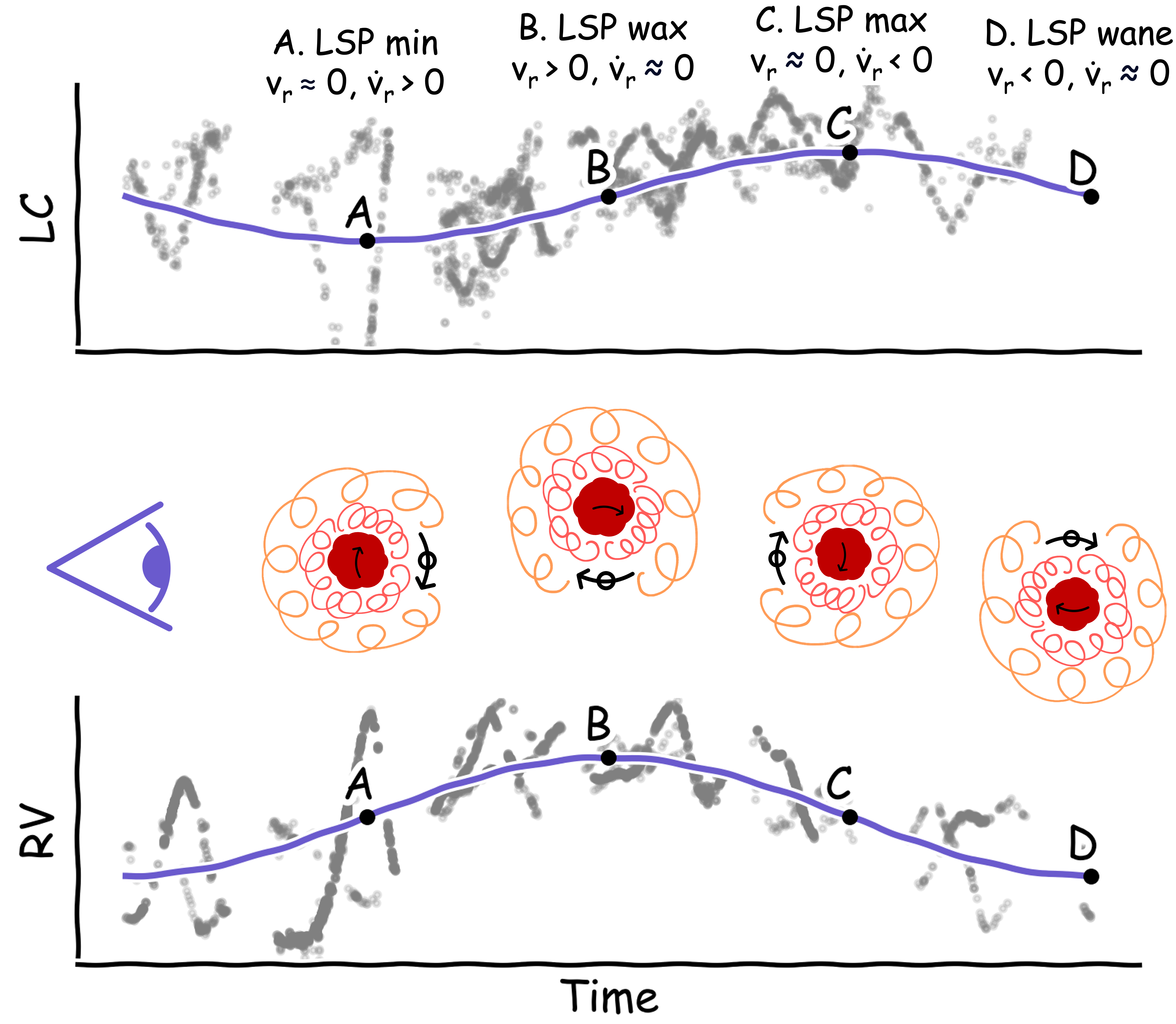}
  \caption{ 
  Schematic of the dust modulation by a companion as discussed in Section~\ref{sec:buddyscenario} over the course of an orbit. 
  The diagram in the center shows 4 orientations of the companion (black circle) as well as Betelgeuse (solid red blob) and some circumstellar gas and dust (curly lines). The arrows represent the motion of the two objects.
  This is compared to the period in the lightcurve (upper panel) and RV modulation (lower panel). 
  The phase-folded data from Figure~\ref{fig:lightcurve_vs_RV} are over-plotted in grey, with 4 highlighted phases labelled as A, B, C, and D. These approximately correspond to (A) the minimum brightness (increasing RV, with the stellar photosphere accelerating away from the observer), (B) increasing brightness (approximate RV maximum), (C) maximum brightness (decreasing RV, star accelerating towards the observer), and (D) decreasing brightness (minimum RV). The Great Dimming of 2020 can be seen in the phase-folded lightcurve near LSP minimum (A). When $\bbud$ is behind Betelgeuse relative to the observer's line of sight as indicated by the RV (configuration A), circumstellar dust is able to obscure the system. When $\bbud$ is in front of Betelgeuse (configuration C), dust unaffected by $\bbud$ is (at least partially) eclipsed by Betelgeuse.}
     \label{fig:schematic}
\end{figure*}

\subsection{Summary of Hypotheses}
Table~\ref{table:summary} provides a quick-look reference for the explanatory power and limitations of each theory discussed in this section.

%
%
\section{Companion-as-LSP in Betelgeuse} 
\label{sec:properties}
In Section \ref{sec:hypotheses}, we argued that the most plausible explanation for the LSP in Betelgeuse is a low-mass companion ($\bbud$) that impacts the dust in its vicinity, corresponding to a brightness increase when the companion is in view, a novel revision of the scenario described by \citet{Soszynski2021}. Figure~\ref{fig:schematic} gives a schematic diagram of the $\alpha$ Ori system proposed in Section~\ref{sec:buddyscenario} over one cycle of the LSP, and illustrates the approximate location of $\bbud$ at various points in the phase-folded lightcurve/RV cycles from Figure~\ref{fig:lightcurve_vs_RV}: 
(A) increasing RV at LSP minimum ($\bbud$ behind Betelgeuse), (B) maximum RV at LSP median ($\bbud$ on the limb moving into view), (C) decreasing RV at LSP maximum ($\bbud$ in view), and (D) minimum RV at LSP median ($\bbud$ on the limb moving out of view). In the remainder of this section, we constrain the orbital properties of the posited $\bbud$, and explore the implications of such a companion. 

\subsection{Properties of Companion}
\label{subsec:prop}
From the RV data we estimate the mass of $\bbud$ via the binary mass function:
\begin{equation}
   f = \frac{M_2^3 \sin^3 i}{(M_1+M_2)^2} = \frac{P\,K^3}{2\pi\,G}
\end{equation}
where $K$ is the velocity amplitude and $P=2169\pm5.3$\,d is the LSP period from the RV observations of \citet{Granzer2022}, and $M_1$ is the present-day mass estimate of $18\pm1M_\odot$ from \citet{Joyce2020}. 
For simplicity, we assume a circular orbit. 
This assumption is further justified given the expected effects of dust reducing eccentricity, investigated primarily in the context of planets in dusty disk environments by, e.g. \citet{Li2019}, \citet{Coleman2022}.
This gives us a lower mass limit of $M_{\rm 2}\sin i = 1.17 \pm 0.07\,M_\odot$. 
This value is an order of magnitude larger than the mass estimates calculated for LSPs in RGBs with RV constraints, which are typically in the $0.1\,M_\odot$ or less range, i.e., closer to brown dwarfs than to dwarf stars \citep{Nicholls2009}. 
In order to cause substantial modulation over the course of an orbit, we assume that a relatively low inclination relative to our line of sight (or a high inclination $i$ relative to the plane perpendicular to our line of sight, as defined conventionally for binary stars and exoplanets) is needed for any interaction with the circumstellar dust to be observable from our vantage point. Therefore, we estimate that the mass of $\bbud$ is unlikely to exceed $1.5\,M_\odot$, or $2\,M_\odot$ at most with a vertically extended cloud.

Using this mass constraint, we calculate the orbital separation assuming a Keplerian orbit: 
\begin{equation}
    P=2\pi\sqrt{\frac{a^3}{G(M_1+M_2)}}
\end{equation}
which recovers an orbital separation of $a = 1850 \pm 70 R_\odot$ assuming the $18\pm1 \Msun$ present-day mass of Betelgeuse estimated by \citet{Joyce2020}. 
Using their stellar radius of $R=764^{+116}_{-62}\,R_\odot$ yields an orbital distance of $a=2.43^{+0.21}_{-0.32}\,R_{*}$. 
This is well outside of the CO2-H2O MOLsphere described by \citet{Montarges2021}, which extends to $\approx1.2R_*$. It also places the L1 Lagrange point at $\approx1380R_\odot$, or $\approx1.8R_*$.  In this orbital configuration, the surface of Betelgeuse is not expected to deform significantly, as the Roche radius for this orbit is $\approx1100R_\odot$, and the equipotential corresponding to $764R_\odot$ is nearly spherical. From the calculated separation, we can estimate a transit probability of $P = R_*/a = 41_{-3}^{+6}\%$, and a corresponding minimal inclination of $i_{\rm min} = 66_{-4}^{^+2\circ}$ for a grazing transit of $\bbud$ (where an inclination of $90^\circ$ means an edge-on orbit). Therefore, even with an extended dust feature that can reach higher inclinations, our assumption of low inclination relative to the observer's line of sight is justified.

It remains unclear from these orbital parameters alone whether $\bbud$ is able to modulate dust. Any irradiation effects will depend on the companion's age and outgoing radiation field and the dust properties, which dynamics alone cannot recover (see further discussion in Section~\ref{subsec:observability}). 
From dynamical arguments, one can calculate the Hill sphere of the companion star, which approximates its sphere of gravitational influence,
\begin{equation}
R_\mathrm{H}\approx{}a\sqrt[3]{\frac{M_2}{M_1+M_2}}.
\end{equation}
This gives $R_\mathrm{H}\approx0.25-0.35a$ within the likely ranges for $\bbud$'s mass, or $60-80\%$ of the radius of Betelgeuse. This is at least a plausible angular cross section to subtend over which one might see modulation. 

\subsection{Observability of Companion} 
\label{subsec:observability}
Even at its highest plausible mass---say 
$M_2\ltapprox2M_\odot$---it would be nearly impossible to detect $\bbud$ close to such a luminous and intrinsically variable star (the average brightness of Betelgeuse hovers near $L\sim10^5\Lsun$) except in some exotic and improbable scenarios (see Section~\ref{subsec:stupidcompanionideas}). 
Even adopting an optimistic 10$\Lsun$ for $\bbud$, the luminosity contrast would still be $\approx10^{-4}$, likely undetectable in integrated light. This challenge is exacerbated by the fact that Betelgeuse's intrinsic variability from convection and pulsations entails $\approx1$mag fluctuations in the V-band (see our Fig.\ref{fig:lightcurve_vs_RV}), with intrinsic fluctuations on the order of more than $\pm\sim10^4\Lsun$, at least 1000 times the intrinsic brightness of $\bbud$. 

Spectral differencing at different phases 
would be another possible avenue for detection, but again, subtle changes in the $\Teff$ of Betelgeuse and changes on the FM timescale would overwhelm variations in spectra unless the companion were substantially hotter than Betelgeuse. 
At its hottest, a 2$\Msun$ star might be spectral class F, corresponding to a $\Teff$ of $\approx6000-7000$K. A companion of this temperature would only just barely be visible on the Wien tail in the UV (a 7000K blackbody outshines a $10^4\times$ brighter, 3600K blackbody only above $1.7\times10^{15}$Hz, or below $\approx1760$\AA{}), and realistically, this is a drastic overestimate of the temperature of $\bbud$.

\subsubsection{Exotic Scenarios}
\label{subsec:stupidcompanionideas}
While the companion's mass and proximity to a massive star might lead one to speculate that it may be a neutron star or other exotic compact body, as far as the authors of this study are aware, no such detections have been made \citep{Posson-Brown-2006,Kashyap-2020}. 
The current lack of X-ray evidence notwithstanding, there are a few reasons not to entirely ignore a neutron-star-as-companion scenario. First, it is more common than not that high-mass stars such as Betelgeuse are born as binary or even triple systems \citep{Sana2012, deMink2013, Toonen2018, Toonen2020, Toonen2022, Offner2023}, many of which will produce runaway stars when the most massive star in the system explodes \citep[e.g.][]{Eldridge2011, Zapartas2021}. Second, the anomalous proper motion of Betelgeuse \citep[e.g.][]{Harper2008} still requires explanation, and a neutron star companion could provide one. In that scenario, the progenitor to a neutron star born at the same time as Betelgeuse must have been more massive than Betelgeuse in order to die earlier in a Supernova explosion. The past SN explosion of Betelgeuse’s companion would have then been energetic enough to provide the necessary “kick” that sent Betelgeuse along its present observed trajectory, with the binary widening but without disrupting the system \citep{Renzo2019}.
Future X-ray observations scheduled at favorable epochs might rule this out conclusively. Such an experiment could also help characterize the presence of a possible corona around a non-compact companion, which might yield insight into the exact mechanism for interaction with any circumstellar dust. 

Using an approximate age for Betelgeuse of $\sim10$Myr, and assuming Betelgeuse and its companion are co-natal, it is possible that $\bbud$ could be a low-mass, pre-main sequence star hosting an accretion or proto-planetary disk. While this could be bright enough to detect, given typical disk lifetimes between $10^5-10^7$ years \citep{Fiorellino2023}, the probability of catching the system in this particular configuration is exceedingly low. The authors are likewise unaware of any existing observational evidence of this scenario.

\subsubsection{Observability Under Normal Conditions}
For spectral differencing or photometry, the closeness of the orbit can be exploited, since the probability of the companion being eclipsed from our vantage point is $P = R_*/a \approx 0.41$. Therefore, an eclipse in the UV passbands would be potentially detectable. Future space-based UV spectroscopic missions such as UVEX \citep{Kulkarni2021} would be well-positioned to investigate this. 
However, one still would have to contend with the length of an eclipse. As for the orbital parameters and a near-circular orbit, the approximate time for $\bbud$ to cross the disk of Betelgeuse is $\approx300$ days, i.e., longer than an observing season, and comparable to the FM and O1 pulsation periods. 

Another possibility is to inspect the shape of evolving asymmetries in sub-milimeter observations of the dusty region around Betelgeuse using an instrument like ALMA. There is indeed substantial asymmetry in the dust observed near the radius of the predicted orbital separation of $\bbud$ \citep[e.g.][]{Richards2013,o'gorman-2017,OGorman2020, Kervella2018}. 

As the sensitivity and achievable luminosity contrast of instruments with powerful coronagraphs and extreme adaptive optics systems designed for planet-hunting and other purposes improve, we are optimistic that in the future such a companion will be detectable, and in the meanwhile we emphasize this system will make for an interesting target for creative observers. 

For future targeted observations, the RV phase can be used to predict where $\bbud$ will be relative to the star over the next LSP cycle and beyond.
Directly observing a transit of $\bbud$ is impossible given Betelgeuse's intrinsic periodic and stochastic variability from convection and pulsations on shorter timescales than the transit. However, the times when $\bbud$ is near the limb could in theory reveal its presence and stellar properties if observed at sufficiently high contrast and spatial resolution. Given the LSP period of 2170 days, the following are the upcoming epochs where $\bbud$ would be near maximum separation: 
\begin{itemize}
    \item Min RV: JD2460651 06/12/2024
    \item Max RV: JD2461736 26/11/2027
    \item Min RV: JD2462821 15/11/2030
    \item Max RV: JD2463906 04/11/2033
\end{itemize}
with syzygy halfway in between.

\begin{table*}
    \centering
    \begin{tabular}{lll}   \hline
         Parameter & Value & Reference\\\hline
         Radius of Betelgeuse  & $764^{+116}_{-62}\,R_\odot$ & \citet{Joyce2020}\\
         Mass of Betelgeuse & $18\pm1\,M_\odot$ & \citet{Joyce2020}\\
         Radius of Betelgeuse  & $764^{+116}_{-62}\,R_\odot$ & \citet{Joyce2020}\\
         Orbital Period  & 2169$\pm$5.3d & \citet{Granzer2022,Jadlovsky2023}\\
         $M\sin{}i$ of $\bbud$ & $1.17 \pm 0.07\,M_\odot$ & This work\\
         Orbital Separation $a$ & $1850 \pm 70 R_\odot$ & This work\\
         D/M/Y of next RV min & 06/12/2024 & This work\\
         D/M/Y of next RV max & 26/11/2027 & This work\\
         D/M/Y of second RV min & 15/11/2030 & This work\\
         D/M/Y of second RV max & 04/11/2033 & This work\\\hline
    \end{tabular}
    \caption{Stellar and orbital parameters recovered for Betelgeuse and $\bbud$.}
    \label{tab:params}
\end{table*}

In the absence of instrumentation breakthroughs enabling extremely high luminosity-contrast direct imaging for stars as bright as Betelgeuse ($L_\odot \sim10^{5}$), we believe the best chance for observability or falsification is with repeated targeted radio-interferometric observations continuously throughout the LSP cycle. These studies should look specifically for patterns in the dust modulation and asymmetry correlated with $\bbud$'s expected orbital separation.

A similar observing strategy was undertaken by \citet{OGorman2015}, who took VLA measurements at 0.7, 1.3, 2.0, 3.5, and 6.1cm at multiple different phase points along the LSP over a 4-year cycle (in addition to single-epoch observations by \citet{Lim1998} 4 years prior). Their Fig.~1 does indicate some time-dependent modulation of an asymmetric circumstellar envelope visible in the 0.7cm interferometric data around $\approx50$mas\footnote{At maximum separation, $\bbud$ will be at $\approx52$mas for a stellar optical photosphere diameter of 43 mas)}, but it is far from conclusive evidence, as an 8-year time span entails measurements during multiple \textit{different} LSP cycles (orbits of $\bbud$).
While we do expect intrinsic diversity in dust structures on a replenishment timescale commensurate with the FM as the FM interacts with the large-scale convection, $\bbud$ would cause a systematic modulation in the circumstellar dust (the proposed mechanism for the LSP itself). 

If an apparent under-density of dust migrates around the star on the LSP cycle, and if that matches the points in the LSP cycle where $\bbud$ should be on the limb versus transiting Betelgeuse versus in opposition, this would lend strong evidence to the existence of $\bbud$ as proposed. Such a finding would also further constrain the system's orbital plane and inclination, the companion's stellar properties, and the mechanism by which $\bbud$ interacts with the dust.

%
%
\section{Conclusions} 
\label{sec:conclusions}

In this study, we have demonstrated that the most likely explanation for the prominent 2170-d periodicity observed in Betelgeuse is a low-mass, binary companion, dubbed $\bbud$, that modulates dust. The proposed orbital picture leads to reduced obfuscation (higher apparent stellar $L$) when $\bbud$ is in transit. 

In Section \ref{sec:LSPs}, we reviewed the variability phenomenon known as Long Secondary Periods (LSPs) present in roughly 30\% of all long-period variable stars and in Betelgeuse specifically. 
In Section \ref{sec:phaseoffset}, we compiled and presented new data (Figure \ref{fig:lightcurve_vs_RV}) comparing Betelgeuse's lightcurve and RV cycle side-by-side and characterized the phase offset between them. We showed that the phase offset present not only in Betelgeuse, but in the majority of Gaia LSP stars, contradicts the physical picture set by the prevailing theory for binarity as the driver for LSPs proposed by \citet{Soszynski2021}. 

In light of the tension introduced by the RV-lightcurve phase offset, we extensively investigated the explanatory power of every hypothesis proposed for LSPs in the literature as applied to Betelgeuse.
Section \ref{sec:hypotheses} covered possibilities ranging from large-scale convection, magnetism, mode interactions, non-radial pulsations, and exotic physics, but only converged on one viable explanation: Betelgeuse's LSP is caused by a low-mass companion, $\bbud$, in a dusty environment.  

In Section \ref{sec:properties}, we calculated physical properties of the companion and the $\alpha$ Orionis orbital system as constrained by existing observations. 
A summary of these parameters is provided in Table~\ref{tab:params}. We discussed why $\bbud$ has not been detected before, commented on exotic physical scenarios that would be theoretically observable, discussed practical limitations on the companion's detectability, and described the advances in method, technique and instrumentation needed to detect it in the future.%

While our work was in the refereeing process, \citet{Macleod2024} independently found astrometric evidence that a companion scenario can explain Betelgeuse's LSP, focusing their efforts on dissecting Betelgeuse's long baseline RV observations. They recovered similar orbital parameters and a slightly lower mass ($M\ltapprox1M_\odot$) due to a smaller inferred RV amplitude.

The confirmation of $\bbud$ would have far-reaching implications for Betelgeuse and its evolutionary future. Betelgeuse is expected to explode as a Type II-Plateau Supernova, and when it does, it could appear nearly as bright as the quarter-moon \citep{Goldberg2020RNAAS}. There is substantial uncertainty in the expected early supernova luminosity evolution due to the unknown status of Betelgeuse's pulsation phase at the time of explosion \citep{Goldberg2020a, Hsu2024} and the properties of the near-surface material 
\citep[e.g.][]{Morozova2017,Morozova2018,Hiramatsu2021b,Moriya2011,Moriya2018,Kozyreva2022,Moriya2023,JacobsonGalan2024}. 

The presence of dense circumstellar material is of reinvigorated interest after the recent Supernova 2023ixf, the closest core-collapse event of the decade, which showed strong signatures of early interaction in an asymmetric circumstellar medium 
\citep[e.g.][]{Hiramatsu2023,JacobsonGalan2023,Hosseinzadeh2023,Smith2023,Teja2023,Zimmerman2024,Li2024,Martinez2024,Singh2024,Moriya2024,Bostroem2024} 
around a pulsating Red Supergiant progenitor \citep[e.g.][]{Soraisam2023,Jencson2023,Kilpatrick2023,Xiang2024}. 
Moreover, supernova remnants are now being observed with unprecedented resolution in the era of JWST \citep[e.g. recent work by][]{Temim2024,Milisavljevic2024}, and the proposed observations of Betelgeuse's surrounding environment, whether or not they support the presence of $\bbud$, will be valuable in constraining the expected level and geometry of circumstellar asymmetries that may imprint on the remnant morphology \citep{Polin2022,Mandal2023,Mandal2024}.
If the existence of $\bbud$ is confirmed and a progenitor identified, one could also predict its own future evolution in the context of Betelgeuse's eventual explosion. Such simulations have recently been carried out in the case of Type Ia SNe with a surviving companion \citep{Bauer2019,Bhat2024,Wong2024}.  

Observations of Betelgeuse have identified photometric variations on a 30 year period consistent with a rotational $v\sin{}i\approx5$km/s \citep{Joyce2020}, which for an inclination angle of 20$^\circ$ recovers an equatorial velocity of 15~km/s (see \citealt{Dupree1987,Gilliland1996,Uitenbroek1998,Kervella2009,Kervella2018}, and review by\citealt{Wheeler2023}). This has led to speculation that Betelgeuse may be a merger product \citep[e.g.][]{Wheeler2017, Chatzopoulos2020,Shiber2024}. This hypothesis was invoked to account for fast envelope rotation which would not survive the spin-down as Betelgeuse ascended up the RSG branch in a single-star evolutionary scenario. 
Although the large-scale convective envelope itself might be the source of this RV signal \citep{Ma2024}, if the apparent RV dipole signal persists at a consistent inclination or at higher ALMA resolution, another possible explanation might be angular momentum transferred from the orbit of $\bbud$, which could be informed by further constraining the stellar mass and properties of $\bbud$.

In this picture, it is not purely a coincidence that the Great Dimming of 2020 \citep[e.g.][]{Dupree2020,Dupree2022,Montarges2021,Drevon2024} occured near LSP Minimum. 
As the opening angle of such an ejection is expected to be large if caused by a large pressure perturbation from convection in the interior \citep{MacLeod2023}, a transiting companion would likely disrupt its path. However, a companion on the other side of Betelgeuse's potential would likely not interact. One may speculate that dusty mass loss could be funneled through the orbit's L3 lagrange point, though dynamical modeling including gas-dust interactions are necessary to quantiatively evaluate this claim.

\acknowledgements
JAG, MJ, and LM contributed equally to this manuscript. We thank the anonymous referee for a constructive report which improved the quality of this manuscript. We acknowledge helpful discussions with Iman Behbehani, Lars Bildsten, Matteo Cantiello, Thavisha Dharmawardena, Zarina Dhillon, Andrea Dupree, Jim Fuller, Margarita Karovska, Shing-Chi Leung, Joseph Long, Jing-ze Ma, Morgan MacLeod, Brian Metzger, Mathieu Renzo, Jamie Tayar, J.~Craig Wheeler, and Chris White. 
We thank the following for their contributions to discussion of observational follow-up: Katie Breivik, Annalisa Calamida, Maria Drout, Christian Johnson, Max Moe, Brendan O'Connor, and Anna O'Grady. We thank John Bourke for typesetting. MJ and LM thank the hospitality of the Flatiron Institute where this project was carried out. 

M.J. gratefully acknowledges funding of MATISSE: \textit{Measuring Ages Through Isochrones, Seismology, and Stellar Evolution}, awarded through the European Commission's Widening Fellowship. This project has received funding from the European Union's Horizon 2020 research and innovation programme. This research was supported by the `SeismoLab' KKP-137523 \'Elvonal grant and by the NKFIH excellence grant TKP2021-NKTA-64 of the Hungarian Research, Development and Innovation Office (NKFIH).  The Flatiron Institute is supported by the Simons Foundation.

We acknowledge with thanks the variable star observations from the AAVSO International Database contributed by observers worldwide and used in this research. This work has made use of data from the European Space Agency (ESA) mission
{\it Gaia} (\url{https://www.cosmos.esa.int/gaia}), processed by the {\it Gaia}
Data Processing and Analysis Consortium (DPAC,
\url{https://www.cosmos.esa.int/web/gaia/dpac/consortium}). This research made use of NASA’s Astrophysics Data System Bibliographic Services, as well as of the SIMBAD database and the cross-match service operated at CDS, Strasbourg, France.

\bibliographystyle{aasjournal}
\singlespace

\bibliography{RSG3D.bib, LSP_ADS.bib, BetelBuddy.bib}

\appendix

\section{Data tables}
In this section we present the following data in numerical format. Table~\ref{tab:rsg_data} lists the brightnesses, distances and periods for the Galactic supergiant stars from Fig.~\ref{fig:PL}. Table~\ref{tab:oglelpv} lists the phase offset values we calculated for the stars presented by \citet{Nicholls2009}, whereas Tables~\ref{tab:gaia1} and \ref{tab:gaia2} contains the phase offset for the subset of \textit{Gaia} LSP candidates that we include in Fig.~\ref{fig:Gaia_LSP}.

\begin{table*}[b]
\centering
\begin{tabular}{lccccccccccc}
\hline
\hline
ID & $K_s$ & d &  $M_K$ & P1 & eP1 & P2 & eP2 & P3 & eP3 & Ref  &  \\
~ & (mag) & (pc) &  (mag) & (d) & (d) & (d) & (d) & (d) & (d) & ~  &  \\
\hline
SS And & 0.971 & $592_{-20}^{+18}$ & $-7.89_{-0.07}^{+0.06}$ & 159 & 17 & ~ & ~ & ~ & ~  & 1 \\
NO Aur & 3.729 & $1183_{-60}^{+51}$ & $-6.64_{-0.11}^{+0.09}$ & 362 & 11 & 38.4 & 0.3 & ~ & ~  & 1 \\
VY CMa & 0.291 & $850_{-157}^{+2259}$ & $-9.36_{-0.44}^{+2.82}$ & 1600 & 190 & ~ & ~ & ~ & ~  & 1 \\
RT Car & 1.500 & $2134_{-99}^{+121}$ & $-10.15_{-0.10}^{+0.12}$ & 201 & 25 & 448 & 146 & ~ & ~  & 1 \\
CL Car & 1.539 & $2421_{-126}^{+148}$ & $-10.38_{-0.12}^{+0.13}$ & 490 & 100 & 229 & 14 & 2600 & 1000  & 1 \\
EV Car & 0.788 & $2424_{-227}^{+262}$ & $-11.13_{-0.21}^{+0.22}$ & 276 & 26 & 820 & 230 & ~ & ~  & 1 \\
IX Car & 1.884 & $2188_{-91}^{+86}$ & $-9.82_{-0.09}^{+0.08}$ & 408 & 50 & 4400 & 2000 & ~ & ~  & 1 \\
TZ Cas & 1.939 & $2335_{-157}^{+147}$ & $-9.90_{-0.15}^{+0.13}$ & 3100 & 1000 & ~ & ~ & ~ & ~  & 1 \\
PZ Cas & 0.781 & $2586_{-254}^{+286}$ & $-11.28_{-0.22}^{+0.23}$ & 850 & 150 & 3195 & 800 & ~ & ~  & 1 \\
ST Cep & 1.644 & $3974_{-315}^{+356}$ & $-11.35_{-0.18}^{+0.19}$ & 3300 & 1000 & ~ & ~ & ~ & ~  & 1 \\
$\mu$ Cep & -1.620 & $1818_{-485}^{+1039}$ & $-12.92_{-0.67}^{+0.98}$ & 860 & 50 & 4400 & 1060 & ~ & ~  & 3 \\
T Cet & -0.808 & $260_{-18}^{+23}$ & $-7.88_{-0.16}^{+0.18}$ & 298 & 3 & 161 & 3 & ~ & ~  & 2 \\
RW Cyg & 0.640 & $1649_{-105}^{+124}$ & $-10.45_{-0.14}^{+0.16}$ & 580 & 80 & ~ & ~ & ~ & ~  & 1 \\
AZ Cyg & 1.288 & $2290_{-115}^{+129}$ & $-10.51_{-0.11}^{+0.12}$ & 495 & 40 & 3350 & 1100 & ~ & ~  & 1 \\
BC Cyg & 0.299 & $1727_{-152}^{+150}$ & $-10.89_{-0.20}^{+0.18}$ & 720 & 40 & ~ & ~ & ~ & ~  & 1 \\
TV Gem & 0.947 & $2344_{-558}^{+1400}$ & $-10.90_{-0.59}^{+1.02}$ & 426 & 45 & 2550 & 680 & ~ & ~  & 1 \\
W Gem & 1.850 & $1799_{-107}^{+145}$ & $-9.42_{-0.13}^{+0.17}$ & 353 & 24 & ~ & ~ & ~ & ~  & 1 \\
BU Gem & 0.806 & $1757_{-306}^{+439}$ & $-10.42_{-0.42}^{+0.48}$ & 2450 & 750 & ~ & ~ & ~ & ~  & 1 \\
$\alpha$ Her & -3.511 & $110_{-14}^{+19}$ & $-8.72_{-0.29}^{+0.34}$ & 124 & 5 & 500 & 50 & 1480 & 200  & 3 \\
W Ind & 3.108 & $1256_{-62}^{+84}$ & $-7.39_{-0.11}^{+0.14}$ & 193 & 15 & ~ & ~ & ~ & ~  & 1 \\
Y Lyn & -0.688 & $350_{-11}^{+16}$ & $-8.41_{-0.07}^{+0.10}$ & 133 & 3 & 1240 & 50 & ~ & ~  & 1 \\
XY Lyr & -0.213 & $415_{-18}^{+21}$ & $-8.30_{-0.10}^{+0.11}$ & 122 & ~ & ~ & ~ & ~ & ~  & 1 \\
$\alpha$ Ori & -4.378 & $168_{-15}^{+28}$ & $-10.51_{-0.20}^{+0.33}$ & 410 & 32 & 200 & 20 & 2100 & 200  & 5 \\
S Per & 1.123 & $2421_{-96}^{+104}$ & $-10.80_{-0.09}^{+0.09}$ & 813 & 60 & ~ & ~ & ~ & ~  & 4 \\
T Per & 2.581 & $2200_{-128}^{+120}$ & $-9.13_{-0.13}^{+0.12}$ & 2500 & 460 & ~ & ~ & ~ & ~  & 1 \\
W Per & 1.568 & $1746_{-90}^{+152}$ & $-9.64_{-0.11}^{+0.18}$ & 500 & 40 & 2900 & 300 & ~ & ~  & 1 \\
RS Per & 1.562 & $2419_{-264}^{+343}$ & $-10.36_{-0.25}^{+0.29}$ & 4200 & 1500 & ~ & ~ & ~ & ~  & 1 \\
SU Per & 1.455 & $2211_{-120}^{+128}$ & $-10.27_{-0.12}^{+0.12}$ & 430 & 70 & 3050 & 1200 & ~ & ~  & 1 \\
XX Per & 1.972 & $2277_{-159}^{+233}$ & $-9.81_{-0.16}^{+0.21}$ & 3150 & 1000 & ~ & ~ & ~ & ~  & 1 \\
BU Per & 2.194 & $2267_{-162}^{+216}$ & $-9.58_{-0.16}^{+0.20}$ & 381 & 30 & 3600 & 1000 & ~ & ~  & 1 \\
FZ Per & 2.482 & $2530_{-140}^{+153}$ & $-9.53_{-0.12}^{+0.13}$ & 368 & 13 & ~ & ~ & ~ & ~  & 1 \\
VX Sgr & -0.122 & $1563_{-92}^{+104}$ & $-11.09_{-0.13}^{+0.14}$ & 754 & 56 & ~ & ~ & ~ & ~  & 4 \\
AH Sco & 0.415 & $1735_{-200}^{+286}$ & $-10.78_{-0.27}^{+0.33}$ & 738 & 78 & ~ & ~ & ~ & ~  & 1 \\
$\alpha$ Sco & -4.100 & $170_{-25}^{+35}$ & $-10.25_{-0.34}^{+0.40}$ & 1650 & 640 & ~ & ~ & ~ & ~  & 3 \\
CE Tau & -0.913 & $708_{-117}^{+174}$ & $-10.16_{--9.77}^{+0.48}$ & 1300 & 100 & ~ & ~ & ~ & ~  & 1 \\
W Tri & 1.091 & $578_{-24}^{+24}$ & $-7.72_{-0.09}^{+0.09}$ & 107 & 6 & 590 & 170 & ~ & ~  & 1 \\
\hline
\end{tabular}
\caption{Red supergiant periods identified by \citet{Kiss2006}. Distance references: 1) \citet{bailer-jones-2021}; 2) \citet{Gaia-EDR3-2021}; 3) \citet{hipparcos-2007}, 4) \citet{Reid-2019}; 5) \citet{Joyce2020}}\label{tab:rsg_data}
\end{table*}

\begin{table}[]
\centering
\begin{tabular}{lcccc}
\hline
\hline
 OGLE ID & $P_{\rm LSP}$ (d) & $\Delta\phi_{RV-I}$ & e$\Delta\phi$ \\
\hline
LMC-LPV-55909 & 783.7 & -1.07 & 0.21 \\
LMC-LPV-55739 & 562.1 & -1.04 & 0.38 \\
LMC-LPV-55861 & 1007 & -1.95 & 0.11 \\
LMC-LPV-55565 & 810.2 & -2.34 & 0.16 \\
LMC-LPV-57329 & 1032 & -2.22 & 0.48 \\
LMC-LPV-56928 & 800 & -1.21 & 0.15  \\
LMC-LPV-56063 & 540 & -3.16 & 0.43  \\
LMC-LPV-57356 & 434.8 & -1.74 & 0.17  \\ 
LMC-LPV-57073 & 470.5 & -1.72 & 0.21  \\
LMC-LPV-58350 & 611.6 & -2.17 & 0.13 \\
LMC-LPV-58399 & 818.7 & 3.90 & 0.13 \\
LMC-LPV-58174 & 640.9 & 3.96 & 0.11 \\
LMC-LPV-59753 & 1048 & 4.36 & 0.29 \\
LMC-LPV-59287 & 257.7 & -3.02 & 0.46 \\
LMC-LPV-60196 & 1139 & -1.93 & 0.35 \\
LMC-LPV-61999 & 404.3 & -3.07 & 0.21 \\
LMC-LPV-60643 & 535.8 & -1.02 & 0.20 \\
LMC-LPV-61577 & 720.5 & -1.75 & 0.15 \\
LMC-LPV-62722 & 715.8 & -2.48 & 0.17 \\
LMC-LPV-63825 & 849.5 & -2.86 & 0.27 \\
\hline
\end{tabular}
\caption{Phase offsets between the RV measurements published by \citet{Nicholls2009} and the OGLE-III \textit{I}-band light curves for stars where those are available. }\label{tab:oglelpv}
\end{table}

\begin{table}
\centering
\begin{tabular}{lcccclcccc}
\hline
\hline
 Gaia DR3 ID & $P_{\rm LSP}$ (d) & $\Delta\phi_{RV-G}$ & e$\Delta\phi$ & Gaia DR3 ID & $P_{\rm LSP}$ (d) & $\Delta\phi_{RV-G}$ & e$\Delta\phi$ \\
\hline
5366556864650293504 & 467.8 & -3.56 & 1.98 & 4518782615952131200 & 413.4 & -1.98 & 0.52 \\
5245519394861788416 & 424.3 & -3.38 & 1.95 & 4591864301979656960 & 384.0 & -1.95 & 0.83 \\
2030155380681518080 & 415.3 & -3.09 & 1.70 & 1958845802441119360 & 445.3 & -1.70 & 0.24 \\
4877073701212505984 & 467.9 & -3.07 & 1.62 & 5823469916709143808 & 433.2 & -1.62 & 0.93 \\
1009476541884049152 & 436.2 & -3.00 & 1.53 & 5404420815320630528 & 561.6 & -1.53 & 0.56 \\
416537855493120896 & 592.8 & -2.54 & 1.48 & 1863461622559594624 & 472.3 & -1.48 & 0.73 \\
5592948576556271360 & 567.5 & -2.54 & 1.47 & 2030200671149815424 & 442.6 & -1.47 & 0.37 \\
5487656658716777984 & 523.9 & -2.52 & 1.43 & 5823227714910793856 & 368.1 & -1.43 & 0.27 \\
4242144837569110656 & 560.2 & -2.52 & 1.36 & 5819332488801340928 & 381.8 & -1.36 & 0.28 \\
5193827058256042752 & 780.7 & -2.47 & 1.35 & 718681245623682176 & 446.5 & -1.35 & 0.35 \\
2940744561181222784 & 510.5 & -2.46 & 1.31 & 5592924490378357632 & 427.0 & -1.31 & 0.46 \\
5393712362310861824 & 378.9 & -2.46 & 1.21 & 2022032910328304128 & 488.3 & -1.21 & 0.15 \\
5365837715327919872 & 294.4 & -2.44 & 1.20 & 2226627353963447552 & 507.6 & -1.20 & 0.73 \\
5193010395993757952 & 585.3 & -2.42 & 1.18 & 6036139001931775360 & 403.9 & -1.18 & 0.39 \\
1863379262264189568 & 339.0 & -2.42 & 1.18 & 413371051550885248 & 436.6 & -1.18 & 0.27 \\
4325121337971096704 & 325.6 & -2.42 & 1.11 & 5842150894131928576 & 367.5 & -1.11 & 0.23 \\
6648842337535201152 & 304.8 & -2.42 & 1.04 & 5615714067889118336 & 605.4 & -1.04 & 0.37 \\
5783448380630645760 & 462.0 & -2.41 & 1.04 & 419317730125131904 & 434.7 & -1.04 & 0.25 \\
525948917547462272 & 770.1 & -2.39 & 1.03 & 5464229059521244928 & 289.8 & -1.03 & 0.41 \\
3417134843227320192 & 814.3 & -2.38 & 0.96 & 4325728161012091904 & 453.3 & -0.96 & 0.44 \\
5870612332427790464 & 328.5 & -2.37 & 0.95 & 2223228905258736128 & 680.9 & -0.95 & 0.94 \\
5793428029206907136 & 466.0 & -2.36 & 0.95 & 5836237308379605248 & 387.0 & -0.95 & 0.32 \\
4216847961220342400 & 353.1 & -2.35 & 0.90 & 2057374688680289408 & 685.6 & -0.90 & 0.91 \\
2162840771749435904 & 701.0 & -2.34 & 0.87 & 563548504362254464 & 547.5 & -0.87 & 0.32 \\
6065282485339565440 & 694.1 & -2.33 & 0.87 & 2111189048343657856 & 498.7 & -0.87 & 0.19 \\
4333627052547566464 & 638.3 & -2.31 & 0.86 & 6076773820652573312 & 397.6 & -0.86 & 0.36 \\
6066743976821731200 & 354.2 & -2.31 & 0.82 & 5779249135267388160 & 462.7 & -0.82 & 0.26 \\
4373808155229106048 & 415.7 & -2.29 & 0.81 & 487598707882598912 & 333.9 & -0.81 & 0.82 \\
5814559886782561152 & 483.4 & -2.28 & 0.81 & 2033040091219687424 & 495.0 & -0.81 & 0.17 \\
1562772473976590976 & 345.3 & -2.27 & 0.78 & 2082877414211373056 & 514.7 & -0.78 & 0.58 \\
4167947800049312128 & 379.9 & -2.27 & 0.75 & 4994896061572855680 & 242.8 & -0.75 & 0.29 \\
4382027799216609280 & 366.1 & -2.26 & 0.73 & 5350758909803889152 & 584.9 & -0.73 & 0.38 \\
5985892282645097728 & 335.5 & -2.24 & 0.73 & 1384487655667958528 & 778.5 & -0.73 & 0.57 \\
5994696415836802304 & 687.5 & -2.21 & 0.69 & 2038497375395267328 & 489.4 & -0.69 & 0.28 \\
5614629674553165568 & 594.9 & -2.20 & 0.68 & 4755154010768073088 & 355.0 & -0.68 & 0.62 \\
2928002527094437504 & 678.3 & -2.20 & 0.67 & 5357213966450101760 & 749.1 & -0.67 & 0.29 \\
5693199740370024576 & 363.9 & -2.19 & 0.66 & 2004024043737361280 & 618.2 & -0.66 & 1.03 \\
6011162495794543104 & 304.5 & -2.18 & 0.65 & 4660125660364324864 & 750.0 & -0.65 & 0.23 \\
5241758687171977216 & 704.5 & -2.17 & 0.64 & 5717734865611449984 & 412.1 & -0.64 & 1.16 \\
6702743863564684800 & 393.9 & -2.16 & 0.63 & 5730565341664785792 & 455.5 & -0.63 & 0.31 \\
2924804941112026240 & 525.0 & -2.14 & 0.63 & 426958575728796032 & 536.2 & -0.63 & 0.14 \\
4487425953063740800 & 683.4 & -2.10 & 0.38 & 5646561553800726784 & 265.4 & -0.38 & 0.29 \\
5914192369148073216 & 460.6 & -2.09 & 0.33 & 4490750562689619584 & 273.9 & -0.33 & 0.35 \\
5846845430839141632 & 482.8 & -2.07 & 0.29 & 4599913964043089408 & 380.3 & -0.29 & 0.19 \\
2192933679121399808 & 674.3 & -2.02 & 0.17 & 5856898196963666816 & 692.4 & -0.17 & 0.29 \\
5970189980950373248 & 482.1 & -2.02 & 0.15 & 2016488970841656320 & 510.8 & -0.15 & 0.29 \\

\hline
\end{tabular}
\caption{Phase offsets between the LPV RV measurements published in the \textit{Gaia} FPR \citep{Trabucchi2023} and the \textit{G}-band \textit{Gaia} DR3 light curves.}\label{tab:gaia1}
\end{table}

\begin{table}
\centering
\begin{tabular}{lcccc}
\hline
\hline
 Gaia DR3 ID & $P_{\rm LSP}$ (d) & $\Delta\phi_{RV-G}$ & e$\Delta\phi$ \\
\hline
2021971062805508224 & 595.7 & -0.09 & 0.30 \\
515032034094986112 & 663.0 & -0.06 & 0.25 \\
2026398010175881856 & 490.4 & 0.29 & 0.61 \\
5243641635146938112 & 350.2 & 0.54 & 0.30 \\
6076305364288485888 & 310.9 & 0.57 & 0.05 \\
2231448029552393472 & 394.8 & -4.85 & 0.15 \\
4688988660900813952 & 370.2 & -4.83 & 0.58 \\
6641986087963535872 & 350.5 & -3.71 & 0.27 \\
5348049300824404864 & 476.4 & -3.58 & 0.44 \\
513141801805888896 & 712.0 & -3.21 & 0.30 \\
2198382038425517696 & 720.6 & -3.21 & 0.30 \\
5834193689811972480 & 466.2 & -3.16 & 0.27 \\
1975711383110161664 & 464.2 & -3.10 & 0.23 \\
5328193048536364672 & 384.6 & -3.08 & 0.40 \\
6649474625438931840 & 289.4 & -3.00 & 0.65 \\
2231049353507901568 & 547.2 & -2.52 & 0.27 \\
5328530461163971968 & 514.8 & -2.41 & 0.82 \\
5829828839147242240 & 435.6 & -2.41 & 0.26 \\
1632704127922550272 & 447.1 & -2.36 & 0.44 \\
2883038617538998144 & 518.9 & -2.35 & 0.26 \\
5646198474439605760 & 539.1 & -2.34 & 0.15 \\
6723996873576976768 & 324.7 & -2.30 & 0.25 \\
1496998382733052928 & 321.9 & -2.28 & 0.30 \\
6076713656747162624 & 623.0 & -2.27 & 0.45 \\
2061975320222333696 & 544.2 & -2.19 & 0.65 \\
5693299903313069312 & 318.1 & -2.11 & 0.28 \\
4601169090926461184 & 384.2 & -1.93 & 1.42 \\
2162696838801402624 & 507.8 & -1.57 & 0.21 \\
5998083014713314944 & 398.2 & -1.56 & 0.44 \\
385255959408720000 & 474.3 & -1.31 & 0.33 \\
2030233136772978304 & 444.6 & -1.14 & 0.55 \\
2895408329510160384 & 564.8 & -1.09 & 0.73 \\
4617724574943762560 & 501.1 & -1.01 & 1.05 \\
2045973161114595584 & 310.2 & -0.96 & 0.24 \\
506098605197562624 & 570.6 & -0.94 & 0.24 \\
6091162171552477440 & 406.6 & -0.91 & 0.20 \\
4569516728144529664 & 406.0 & -0.89 & 0.35 \\
4487550915135532288 & 411.8 & -0.81 & 0.19 \\
4613334534251566848 & 534.7 & -0.80 & 0.28 \\
513591398982018944 & 735.0 & -0.75 & 0.30 \\
2014407045572567424 & 354.5 & -0.68 & 0.46 \\
186324619130226048 & 575.0 & -0.67 & 1.66 \\
505958902802490752 & 589.8 & -0.67 & 0.60 \\
5638830642723891328 & 655.7 & -0.65 & 0.12 \\
\hline
\end{tabular}
\caption{\textit{Gaia} phase offsets, continued. }\label{tab:gaia2}
\end{table}

\end{document}